\documentclass{article}
\usepackage[preprint,nonatbib]{neurips_2020}
\usepackage[utf8]{inputenc} 
\usepackage[T1]{fontenc}    
\usepackage{hyperref}       
\usepackage{url}            
\usepackage{booktabs}       
\usepackage{amsfonts}       
\usepackage{nicefrac}       
\usepackage{microtype}      
\usepackage{xcolor}
\usepackage[linesnumbered,ruled,vlined]{algorithm2e}
\usepackage{tikz} 
\usepackage{graphicx}
\usepackage{amsmath}
\usepackage{amsthm}
\usepackage{euler}
\usepackage{wrapfig}

\newtheorem{lem}{Lemma}

\usetikzlibrary{shapes, arrows, calc, arrows.meta, fit, positioning}
\tikzset{  
  -Latex,auto,node distance =0.3 cm and 1.3 cm, thick,
  state/.style ={circle, draw, minimum width = 0.9 cm}, 
  var/.style ={circle, draw, minimum width = 0.9 cm}, 
  factor/.style ={regular polygon,regular polygon sides=4, fill=black, draw, minimum width = 0.6 cm}, 
}  

\begin{document}

\title{CRISP: A Probabilistic Model for Individual-Level COVID-19 Infection Risk Estimation Based on Contact Data}

\author{Ralf Herbrich\thanks{The ordering of authors is alphabetical. All authors contributed equally to the paper.} \\
Zalando \\
Berlin, Germany \\
\texttt{rherbrich@gmail.com} \\ 
\And
Rajeev Rastogi \\
Amazon \\
Bangalore, India \\
\texttt{rastogi@amazon.com} \\ 
\And
Roland Vollgraf \\
Zalando \\
Berlin, Germany \\
\texttt{roland.vollgraf@gmail.com} \\
}

\maketitle

\begin{abstract}
  We present CRISP ({\bf C}OVID-19 {\bf RI}sk {\bf S}core {\bf P}rediction), a probabilistic graphical model for COVID-19 infection spread through a population based on the SEIR model where we assume access to (1) mutual contacts between pairs of individuals across time across various channels (e.g., Bluetooth contact traces), as well as (2) test outcomes at given times for infection, exposure and immunity tests. Our micro-level model keeps track of the infection state for each individual at every point in time, ranging from susceptible, exposed, infectious to recovered. We develop both a Monte Carlo EM as well as a message passing algorithm to infer contact-channel specific infection transmission probabilities. Our Monte Carlo algorithm uses Gibbs sampling to draw samples of the latent infection status of each individual over the entire time period of analysis, given the latent infection status of all contacts and test outcome data. Experimental results with simulated data demonstrate our CRISP model can be parametrized by the reproduction factor $R_0$ and exhibits population-level infectiousness and recovery time series similar to those of the classical SEIR model. However, due to the individual contact data, this model allows fine grained control and inference for a wide range of COVID-19 mitigation and suppression policy measures. Moreover, the block-Gibbs sampling algorithm is able to support efficient testing in a test-trace-isolate approach to contain COVID-19 infection spread. To the best of our knowledge, this is the first model with efficient inference for COVID-19 infection spread based on individual-level contact data; most epidemic models are macro-level models that reason over entire populations. The implementation of CRISP is available in Python and C++ at \href{https://github.com/zalandoresearch/CRISP}{https://github.com/zalandoresearch/CRISP}. 
\end{abstract}

\section{Introduction}
The COVID-19 pandemic has spread rapidly around the world, with the number of infections and deaths steadily growing. Most governments around the world have been completely unprepared to deal with the COVID-19 outbreak, which UN Secretary-General Antonio Guterres has referred to as humanity’s worst crisis since World War II. While governments around the world had plans in place in the event of a pandemic, the peculiarities of COVID-19 (e.g., delayed onset of symptoms, asymptomatic transmission) have challenged these preparations. Governments have reacted by implementing measures such as nationwide lock-downs, that require people to stay inside their homes, enforcing social distancing and therefore breaking the COVID-19 infection chain. However, a blunt mechanism such as a lock-down (over an extended period) can cause severe damage to the economy, and so, there is a need to find alternative measures to slow down or stop the spread without incremental effects in other areas of society. These alternatives have to be built in a solid foundation such as widespread testing and the isolation of infected (or potentially infected) individuals via contact-tracing.

Contact tracing technologies \cite{TraceTogether,AarogyaSetu} have shown promise in tracking the spread of the disease across the population. These mobile apps capture social contact information between users such as contact duration, distance, etc. using Bluetooth signals on devices. The fine-grained contact data of individuals collected by the apps can enable:
\begin{itemize}
    \item {\it Individual risk score prediction.} The contact data, combined with information about COVID-19 positive test cases, can be used to predict the likelihood of infection for each individual. The individual risk scores can be leveraged by governments and organizations to prioritize testing as well as to identify individuals that need to enter isolation/quarantine. 
    \item {\it Hotspot detection.} Tracing technologies can help authorities identify areas with a high density of contacts and/or individuals with high infection risk. This can allow policymakers to make more effective decisions, for example, by imposing highly restrictive measures such as lock-downs, shelter-at-home, or school closures only in COVID-19 hotspots while allowing activities to remain closer to normal in unaffected areas.
    \item {\it Insights about infection spread.} Contact tracing can provide insights into the relative importance of different modalities of disease transmission (e.g., through intermediate surfaces vs individual contact), risk of infection transmission based on contact characteristics such as duration and distance, most likely locations (e.g., schools, work, malls) for the spread of disease, and "super spreaders" who come in close proximity with a large number of individuals and so must be frequently tested for infection. 
\end{itemize}

To achieve the above-mentioned benefits, we need to devise new models and inference algorithms for analyzing contact tracing data. This is because existing epidemics models \cite{Chakrabarti2008,Mieghem2009,Mieghem2014,Ferguson2005,Ferguson2006} focus on estimating population-level statistics such as percentage of the population infected, number of days for the epidemic to peak, etc. as opposed to the infection state of each individual in the population. Other models \cite{Myers2010,Warriyar2020} that use ML-based inference techniques assume complete knowledge of the infection state of each individual at each time instant. However, in the COVID-19 scenario, (1) the infection status of individuals is not known until they are tested, and (2) infectious time of individuals are unknown since individuals may infect others while asymptomatic. Finally, governments are using contact tracing data \cite{TraceTogether,AarogyaSetu} to identify and test individuals who have come in direct contact with COVID-19 positive test cases. However, the fact that asymptomatic individuals may have infected a large number of people prior to displaying symptoms and being tested, delays the detection of these newly infected individuals by only using contact tracing apps. 

Our main contributions can be summarized as follows:
\begin{itemize}
    \item We propose CRISP ({\bf C}OVID-19 {\bf RI}sk {\bf S}core {\bf P}rediction), a probabilistic graphical model for COVID-19 infection spread through diverse contacts channels between individuals. Our model uses latent variables to represent the epidemiological states of individuals based on the SEIR model \cite{May1991} at different points in time, and captures both the transitions between states as well as test outcomes. 
    \item We develop both a Monte Carlo EM and message passing algorithm to infer infection transmission probabilities across a range of contact channels. Our Monte Carlo algorithm uses block-Gibbs sampling to draw samples of the latent infection status of each individual over the entire time period, given data about contacts and test results. 
    \item We provide implementation details to accelerate  the block-Gibbs sampling, the message passing and the forward sampling algorithm. A Python and C++ implementation of CRISP is available at \href{https://github.com/zalandoresearch/CRISP}{https://github.com/zalandoresearch/CRISP}.
    \item We conduct experiments with simulated data which demonstrate that our CRISP model can be parametrized by the reproduction factor $R_0$ and exhibits population-level infectiousness and recovery time series similar to those of the classical SEIR model. However, due to the individual contact data, this model allows fine grained control and inference for a wide range of COVID-19 mitigation and suppression policy measures. Furthermore, we show that a testing-and-quarantining policy based on infection risk scores computed by the CRISP algorithm is able to mitigate COVID-19 infection spread while quarantining fewer individuals compared to other policies based on contact-tracing and symptom-based testing.  
\end{itemize}

To the best of our knowledge, this is the first comprehensive model for COVID-19 infection spread that (1) captures the infection states of individuals and transitions between them using the SEIR model, and (2) leverages contact tracing and test outcome data to infer model parameters such as contact-channel specific infection rates using scalable and computationally efficient inference algorithms.

\section{Related Work}
We classify related work into four broad categories: epidemic models, Machine Learning (ML) based inference of model parameters, influence maximization in social networks, and contact tracing apps. 

\subsection{Epidemic Models}

In recent years, there has been research on modeling individual dynamics of epidemics \cite{Chakrabarti2008,Mieghem2009,Mieghem2014}. However, this work typically resorts to mean-field theory to model virus spread over a network, and thus does not characterize the dynamic infectious state of each individual over time. 

Ferguson et al. \cite{Ferguson2005,Ferguson2006} use a compartmental transmission model to simulate the spread of influenza across a population, and analyze interventions such as antiviral prophylaxis and social distancing to halt a pandemic. The authors use a stochastic model of individuals co-located in households that are randomly distributed across a geographical region, and infection risk from 3 sources – household, place and random contacts in the community. The infection transmission rates for the 3 sources and recovery rates are based on analysis of historical data. In contrast, we leverage real individual contact tracing data and outcomes of tests on individuals to infer the infection transmission rate for each contact and the likelihood of infection for each individual. 

Lorch et al. \cite{Lorch2020} propose a spatiotemporal epidemic model that uses marked temporal processes to represent the epidemiological condition of each individual (based on a variation of the SEIR compartment models), individual mobility patterns, test outcomes, and testing and contact tracing strategies. The authors design an efficient sampling algorithm for the model using Monte Carlo roll-outs that is able to predict the spread of COVID-19 under different testing \& tracing strategies, social distancing measures, and business restrictions, given contact histories of individuals. They use Bayesian optimization techniques to infer model parameters (e.g. infection transmission rate) that minimize the difference between the real positive COVID-19 cases and those in the Monte-Carlo simulations. In addition, they demonstrate the efficacy of their model using real COVID-19 data and mobility patterns of Tübingen, Germany. Our Monte Carlo EM inference algorithm for model parameters is computationally much more efficient than the Bayesian optimization techniques employed in \cite{Lorch2020}.

\subsection{Machine Learning-based Inference}

In \cite{Myers2010}, the authors consider the problem of inferring latent social networks based on network diffusion or disease propagation data. Given the times when nodes become infected, but not who infected them, the authors identify the optimal network that best explains the observed data. The authors present a maximum likelihood approach based on convex optimization with a $L_1$-like penalty term (that encourages sparsity) to estimate the conditional probability of infection transmission between every node pair. A key difference from our work is that \cite{Myers2010} assumes complete knowledge of infected nodes and infection times. In contrast, in the COVID-19 scenario, (1) the infection status of nodes is not known until they are tested, and (2) infection times of nodes are unknown since nodes may not show symptoms even though they are infected (and infecting others). 

Warriyar et al. \cite{Warriyar2020} introduce a novel R statistical software package EpiILM for simulating infectious disease spread, and carrying out Bayesian MCMC-based statistical inference for spatial and/or (contact) network-based models in the Deardon et al. \cite{Deardon2010} individual-level modelling framework. In individual-level models (ILMs), the epidemiological state of each individual (e.g., susceptible or infected) is assumed to be perfectly known at each time instant, which makes it relatively straightforward to estimate model parameters such as infection transmission probabilities (as a function of covariates) using maximum likelihood estimation or Bayesian inference using Metropolis-Hastings MCMC. However, in the COVID-19  scenario, epidemiological states of individuals are hidden until they are tested, and this complicates Bayesian inference in our probabilistic model setting. 

\subsection{Influence Maximization in Social Networks}

The {\it Influence Maximization} problem aims to select $k$ users in a social network that maximize influence spread, and was first modeled as an algorithmic problem by Kempe et al. \cite{Kempe2003}. \cite{Yuchen2018} presents a comprehensive survey of different diffusion models that capture the information diffusion process and approximation algorithms to maximize influence. The papers assume that diffusion model parameters such as influence probabilities are given and focus on selecting $k$ users to maximize influence spread. In contrast, our paper focuses on the problem of estimating model parameters related to infection transmission probabilities for each contact, given social contact information between users and COVID-19 test results for users. 

\cite{Mathioudakis2011} addresses the problem of finding the "backbone" of an influence network. It employs network sparsification to preserve only the links that play an important role in the propagation of information. \cite{Goyal2010} considers the problem of estimating influence probabilities between users in a social graph. Given a social graph and a log of actions by users, the Maximum Likelihood Estimator (MLE) of influence probability of node $u$ on node $v$ is simply the fraction of actions performed by $u$ that are also performed by $v$. Unlike \cite{Goyal2010}, in our setting, the infection status and times of nodes are latent, and need to be inferred by our algorithms. 

\subsection{Contact Tracing Apps}

To combat the spread of COVID-19, governments have launched contact tracing apps\cite {TraceTogether,AarogyaSetu} that use Bluetooth signals on mobile phones to track contacts between users. Users who have come in direct contact with COVID-19 positive test cases are considered to be at high risk of infection, and subject to tests and quarantine actions. However, a key problem with this approach is that COVID-19 infected users are typically tested only after they show symptoms, and typically, infected users show symptoms 5-6 days post infection. These asymptomatic users may have infected a large number of users over multiple hops prior to displaying symptoms and being tested. This delays detection of infected users using contact tracing apps, and  limits their effectiveness to proactively test and isolate infected users to contain the spread of COVID-19. In contrast, our probabilistic modeling algorithm CRISP predicts the likelihood of a user getting infected with COVID-19 through a chain of social contacts involving asymptomatic users, and identifies infected users early, even though they may be multiple hops from a user who has tested positive for COVID-19 and even before they begin to show symptoms. Our inference algorithm also learns infection transmission probabilities for each contact channel.

\section{CRISP Infection Spread Model}
\begin{figure}
\centering
\begin{minipage}[t]{0.4\textwidth}
\begin{tikzpicture}  
    \node[state] (z11) at (0,0) {$z_{1,1}$}; 
    \node[state] (z12) [right=of z11] {$z_{1,2}$};  
    \node[state] (z13) [right=of z12] {$z_{1,3}$};  
    \node[state] (z14) [right=of z13] {$z_{1,4}$};  

    \node[state] (z21) at (0,-2.5) {$z_{2,1}$}; 
    \node[state] (z22) [right=of z21] {$z_{2,2}$};  
    \node[state] (z23) [right=of z22] {$z_{2,3}$};  
    \node[state] (z24) [right=of z23] {$z_{2,4}$};  

    \node[state] (z31) at (0,-5) {$z_{3,1}$}; 
    \node[state] (z32) [right=of z31] {$z_{3,2}$};  
    \node[state] (z33) [right=of z32] {$z_{3,3}$};  
    \node[state] (z34) [right=of z33] {$z_{3,4}$};  

    \node[state] (o34) [above=of z34] {$o_{3,4}$};  

    \path (z11) edge (z12); 
    \path (z11) edge[bend right=30] (z13); 
    \path (z11) edge[bend left=30] (z14); 
    \path (z12) edge (z13); 
    \path (z12) edge[bend left=30] (z14); 
    \path (z13) edge (z14);     

    \path (z21) edge (z22); 
    \path (z21) edge[bend right=30] (z23); 
    \path (z21) edge[bend left=30] (z24); 
    \path (z22) edge (z23); 
    \path (z22) edge[bend left=30] (z24); 
    \path (z23) edge (z24);     

    \path (z31) edge (z32); 
    \path (z31) edge[bend right=30] (z33); 
    \path (z31) edge[bend left=30] (z34); 
    \path (z32) edge (z33); 
    \path (z32) edge[bend left=30] (z34); 
    \path (z33) edge (z34);     

    \path (z12) edge (z23); 
    \path (z12) edge (z33); 
    \path (z22) edge (z13); 
    \path (z32) edge (z13); 

    \path (z34) edge (o34); 

\end{tikzpicture} 
\end{minipage} 
\hspace{3cm}
\begin{minipage}[b]{0.3\textwidth}
\caption{Graphical model of the CRISP contact infection spread model for 3 people over 4 time steps where individual $u=1$ meets both individual $u=2$ and $u=3$ at time $t=2$ and one test outcome of individual $u=3$ at time $t=4$. Note that this model has no cycles as we assume the infection status $z_{u,t}$ only depends on variables $z_{v,t^\prime}$ {\it before} time step $t$, $t^\prime < t$. However, due to the "memory" that the state $z_{u,t}=E$ and $z_{u,t}=I$ have, we require edges into the entire past of an infection trace. \label{fig:graphical_model}}
\end{minipage}
\end{figure}
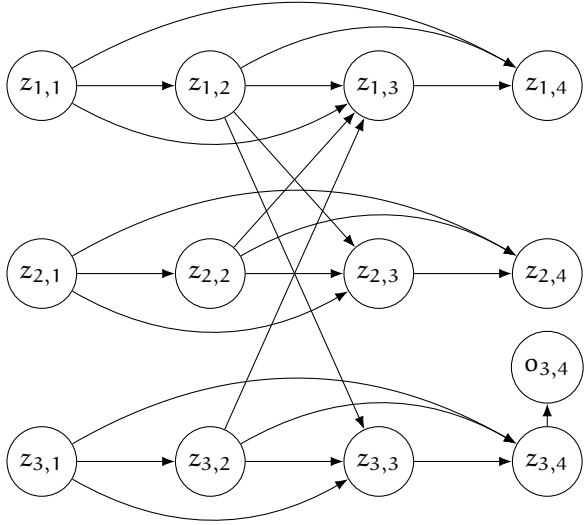

Our CRISP model is an SEIR model at the level of every individual (see \cite{May1991} for an introduction). Note that we consider discrete time steps $t$ implicitly assumed to be at the level of single days. We assume that we are given the following two datasets for a given set $\mathcal{S}$ of individuals:

\begin{itemize}
    \item $\mathcal{D}_{\mathrm{contact}} = \{(u_i,v_i,t_i,\mathbf{x}_i)\}_{i=1}^N \subseteq \mathcal{S} \times \mathcal{S} \times \mathbb{N} \times \mathbb{N}^J$ of $N$ quadruples of a pair of two individuals $(u_i,v_i)$ who have met at time $t_i$ with specific features $\mathbf{x}_i$. Here we assume that the feature vector $\mathbf{x}_i$ describes the overall contact between $u_i$ and $v_i$ via the number $x_{i,j}$ of mutual contacts over channel $j$ (e.g., Bluetooth encounters, queuing together, sharing public transportation). We assume $\mathcal{D}_{\mathrm{contact}}$ to be symmetric so that $(u,v,t,\mathbf{x})\in\mathcal{D}_{\mathrm{contact}} \leftrightarrow (v,u,t,\mathbf{x})\in\mathcal{D}_{\mathrm{contact}}$.
    \item $\mathcal{D}_{\mathrm{test}} :=\{ (u_i,t_i,o_i) \}_{i=1}^K \subseteq \mathcal{S} \times \mathbb{N} \times \{0,1\}$ of $K$ triplets of individual $u_i$ taking a test at time $t_i$ with the test outcome $o_i$ where $o_i=0$ indicates a negative test outcome. 
\end{itemize}

In the following we present two equivalent SEIR models. The first one is a pure four-state SEIR model which has a separate model for the state transitions from $S$ to $E$, $E$ to $I$, and $I$ to $R$. The subsequent subsection present an equivalent model that has the advantage that it is fully Markovian in the sense that a latent variable at time $t+1$ only depends on latent variables at time $t$.

\subsection{Four-State CRISP Model}
\label{subsec:four-state-CRISP-model}
We model the $T$ discrete time steps of infection status for each individual. Our model has $|\mathcal{S}| \times T$ many latent variables $\mathcal{Z}:=\{z_{u,t}\}_{u\in\mathcal{S},t=1,\ldots,T}\in\{S,E,I,R\}^{|\mathcal{S}| \times T}$ that represent the four stages of infection\footnote{Note that we are assuming that recovered individuals are immune until time step $T$.}:

\begin{itemize}
    \item $z_{u,t}=S$: individual $u$ has not been infected and is susceptible,
    \item $z_{u,t}=E$: individual $u$ is infected but not contagious,
    \item $z_{u,t}=I$: individual $u$ is infected and is contagious,
    \item $z_{u,t}=R$: individual $u$ has recovered and is not contagious. 
\end{itemize}

Let us use the notation $\mathcal{Z}_{u,t}:=\{z_{u,1},\ldots,z_{u,t}\}$ to denote the set of latent states $z_{u,t}$ of individual $u$ up to and including time $t$ and $\mathcal{Z}_t := \bigcup_u \mathcal{Z}_{u,t}$. In addition to the latent infection status of all individuals at each time step, we also model $K$ variables $o_{u,t}\in\{0,1\}$ for the test outcomes in $\mathcal{D}_{\mathrm{test}}$, that is, $\mathcal{O}:=\{o_{u,t} : (u,t,o_{u,t}) \in \mathcal{D}_{\mathrm{test}}\}$. Then, our graphical model $\mathcal{G}:=(\mathcal{V},\mathcal{E})$ between the variables $\mathcal{V}:=\mathcal{Z} \bigcup \mathcal{O}$ has the following edges:
\begin{enumerate}
    \item $\mathcal{E}_\mathrm{time} = \bigcup_u \mathcal{E}_u$ and $\mathcal{E}_u := \{(z_{u,t},z_{u,t^\prime})_{t<t^\prime}\}$. All edges between the latent infection states of a single individual $u$. The edges $\mathcal{E}_u$ will be used to describe the probability of $P(z_{u,t}|z_{u,t-1},\ldots,z_{u,1})$ and describe the full time series of being susceptible, exposed, infectious and then recovered.
    \item $\mathcal{E}_\mathrm{contact} := \bigcup_{(u,v,t,\mathbf{x}) \in \mathcal{D}_{\mathrm{contact}}} \{(z_{u,t},z_{v,t+1})\}$. All edges between two individuals $u$ and $v$ who had a contact at time $t$.
    \item $\mathcal{E}_\mathrm{test} := \bigcup_{(u,t,o) \in \mathcal{D}_{\mathrm{test}}} \{(z_{u,t},o_{u,t})\}$. All edges between a test outcome at time $t$ and the corresponding infection status $z_{u,t}$ of individual $u$. These edges will be used to describe the probabilities $P(o_{u,t}|z_{u,t})$ of a test outcome given the infection status of $u$ at that same time.
\end{enumerate}
The full edge set $\mathcal{E}$ is the union of these three edge types: $\mathcal{E} = \mathcal{E}_\mathrm{time} \cup \mathcal{E}_\mathrm{contact} \cup \mathcal{E}_\mathrm{test}$. Figure \ref{fig:graphical_model} shows an example graphical model with these three edge components. 

In order to define the joint probability distribution, note that all edges $\mathcal{E}_\mathrm{time}$ and $\mathcal{E}_\mathrm{contact}$ are pointing forward in time. Thus, all $\{z_{u,t+1}\}_{u\in\mathcal{S}}$ are conditionally independent of each other given all the past states $\mathcal{Z}_t$. Also, all edges $\mathcal{E}_\mathrm{test}$ have the property that a test outcome of individual $u$ at time $t$ only depends on the infection status of $u$ at $t$, $z_{u,t}$. The joint probability distribution is given by
\begin{eqnarray}
P\left(\mathcal{Z}_T,\mathcal{O}\right) & = & P\left(O|\mathcal{Z}_T\right)\cdot P\left(\mathcal{Z}_T\right) \,, \label{eq:joint_probability} \\
P\left(\mathcal{O}|\mathcal{Z}_T\right) & = & \prod_u \prod_{(t,o)\in\mathcal{T}_u} P(o|z_{u,t}) \,, \\
P\left(\mathcal{Z}_T\right) & = & \prod_t \prod_u P(z_{u,t+1}|\mathcal{Z}_t) \,,
\end{eqnarray}
where  $\mathcal{T}_u=\left\{(t,o):(u,t,o)\in\mathcal{D_\textrm{test}}\right\}$. Since we are using an SEIR model, the only non-zero probabilities $P(z_{u,t+1}|\mathcal{Z}_t)$ are the transitions $S \rightarrow S$, $S \rightarrow E$, $E \rightarrow E$, $E \rightarrow I$, $I \rightarrow I$, $I \rightarrow R$, $R \rightarrow R$. 
\begin{equation}
  P(z_{u,t+1}|\mathcal{Z}_t)  = \begin{cases}
    f(u,t,\mathcal{Z}_t) & \mbox{if } z_{u,t}=S \wedge z_{u,t+1}=S\\
    1-f(u,t,\mathcal{Z}_t)  & \mbox{if } z_{u,t}=S \wedge z_{u,t+1}=E\\
    1-g(u,t,\mathcal{Z}_{u,t})  & \mbox{if } z_{u,t}=E \wedge z_{u,t+1}=E\\
    g(u,t,\mathcal{Z}_{u,t})  & \mbox{if } z_{u,t}=E \wedge z_{u,t+1}=I\\
    1-h(u,t,\mathcal{Z}_{u,t})  & \mbox{if } z_{u,t}=I \wedge z_{u,t+1}=I\\
    h(u,t,\mathcal{Z}_{u,t})  & \mbox{if } z_{u,t}=I \wedge z_{u,t+1}=R\\
    1  & \mbox{if } z_{u,t}=R \wedge z_{u,t+1}=R\\
    0  & \mbox{otherwise}
    \end{cases}\,. \label{eq:conditional_model}
\end{equation}

\paragraph{Infection Model} In order to define $f$, we assume that an infection occurs from exogenous influences with a fixed probability $p_0 \in [0,1]$ or with probability of $p_j\in[0,1]$ for every instance of a contact through the contact channel $j$ if the contact was already in the state $I$. Thus, the probability that no infection occurred at time $t$ equals
\begin{equation}
  f(u,t,\mathcal{Z}_t) = (1-p_0) \cdot \displaystyle\prod_{(v,u,t,\mathbf{x}) \in \mathcal{D}_{\mathrm{contact}}: z_{v,t}=I} \prod_{j=1}^J (1-p_j)^{x_j} \,. \label{eq:f_function}
\end{equation}

\paragraph{Infection Status Model} In order to define $g$ and $h$, let us assume we have a point density function $q_E: \mathbb{N}^+ \mapsto [0,1]$ and $q_I: \mathbb{N}^+ \mapsto [0,1]$ for the probability $q_E(d_E)$ that the exposure ($z_{u,t}=E$) lasts for $d_E$ time steps (and similarly for the duration of the infectiousness). Examples of functions $q_E$ and $q_I$ are the probability mass functions of the binomial, negative-binominal or geometric distributions. However, for the case of COVID-19, we will use discrete probabilities established from analysis of the population in \cite{Backer2020} and \cite{Woelfel2020}. Moreover, let
\begin{equation}
  \pi(n;q) = \frac{q(n)}{1-\sum_{i=1}^{n-1}q(i)} = \frac{P(d=n)}{P(d\geq n)} = P(d=n|d\geq n) \,,
\end{equation}
be the conditional probability (according to $q$) that the duration is exactly $n$ time steps given that the duration is at least $n$ time steps. Then,
\begin{eqnarray}
  g(u,t,\mathcal{Z}_{u,t}) & = & \pi\left(t-\max_{t^\prime \leq t}\{t^\prime : z_{u,t^\prime}=S\};q_E\right) \label{eq:g_function} \\
  h(u,t,\mathcal{Z}_{u,t}) & = & \pi\left(t-\max_{t^\prime \leq t}\{t^\prime : z_{u,t^\prime}=E\};q_I\right) \label{eq:h_function} 
\end{eqnarray}
Note that the first argument to both $g$ and $h$ is the number of $E$ and $I$ states up to and including time $t$ in the state sequence $\mathcal{Z}_{u,t}$.

\paragraph{Test Outcome Model}
Finally, we need to define the probability of a test outcome $o$ given the infection status $z_{u,t}$ of individual $u$ at time $t$. Since there are two types of mistakes of a test, we use
\begin{equation}
    P\left(o|z_{u,t}\right) =  
    \begin{cases}
    \alpha & \textrm{if}\ z_{u,t}=I \wedge o=0 \\
    1-\alpha & \textrm{if}\ z_{u,t}=I \wedge o=1 \\
    1-\beta & \textrm{if}\ z_{u,t}\in\left\{ S,E,R\right\} \wedge o=0 \\
    \beta & \textrm{if}\ z_{u,t}\in\left\{ S,E,R\right\} \wedge o=1  
    \end{cases}\,. \label{eq:test_outcome}
\end{equation}
We assume both $0<\alpha\ll 1$ and $0 < \beta \ll 1$. It is easy to implement more sophisticated test accuracy models here, in particular to distinguish between different infection states. Also, we can easily model $\alpha$ and $\beta$ which are dependent on how many days an individual has been in state $I$; this change would not affect the block-Gibbs sampling scheme in Subsection \ref{subsec:efficient-block-Gibbs-sampling} in an adverse way.  

\paragraph{Prior Model} In order to complete the description of the full probabilistic model, we have to specify $P(\mathcal{Z}_1)$. For simplicity, we assume these probabilities to be a delta-peak at state $S$, that is, $P(z_{u,1}=S)=1$ for all $u\in \mathcal{S}$. 

\subsection{Markovian CRISP Model}
\label{subsec:Markovian}
\begin{figure}
\centering
\begin{minipage}[t]{0.4\textwidth}
\begin{tikzpicture}  
    \node[state] (y11) at (0,0) {$y_{1,1}$}; 
    \node[state] (y12) [right=of y11] {$y_{1,2}$};  
    \node[state] (y13) [right=of y12] {$y_{1,3}$};  
    \node[state] (y14) [right=of y13] {$y_{1,4}$};  

    \node[state] (y21) at (0,-2.5) {$y_{2,1}$}; 
    \node[state] (y22) [right=of y21] {$y_{2,2}$};  
    \node[state] (y23) [right=of y22] {$y_{2,3}$};  
    \node[state] (y24) [right=of y23] {$y_{2,4}$};  

    \node[state] (y31) at (0,-5) {$y_{3,1}$}; 
    \node[state] (y32) [right=of y31] {$y_{3,2}$};  
    \node[state] (y33) [right=of y32] {$y_{3,3}$};  
    \node[state] (y34) [right=of y33] {$y_{3,4}$};  

    \node[state] (o34) [above=of y34] {$o_{3,4}$};  

    \path (y11) edge (y12); 
    \path (y12) edge (y13); 
    \path (y13) edge (y14);     

    \path (y21) edge (y22); 
    \path (y22) edge (y23); 
    \path (y23) edge (y24);     

    \path (y31) edge (y32); 
    \path (y32) edge (y33); 
    \path (y33) edge (y34);     

    \path (y12) edge (y23); 
    \path (y12) edge (y33); 
    \path (y22) edge (y13); 
    \path (y32) edge (y13); 

    \path (y34) edge (o34); 

\end{tikzpicture} 
\end{minipage} 
\hspace{3cm}
\begin{minipage}[b]{0.3\textwidth}
\caption{Graphical model of the Markovian CRISP contact infection spread model for 3 people over 4 time steps where individual $u=1$ meets both individual $u=2$ and $u=3$ at time $t=2$ and one test outcome of individual $u=3$ at time $t=4$. Note that this model is equivalent to the model in Figure~\ref{fig:graphical_model} and also has no cycles as the infection status $y_{u,t}$ only depends on variables $y_{v,t^\prime}$ {\it before} time step $t$, $t^\prime < t$. \label{fig:markovian_graphical_model}}
\end{minipage}
\end{figure}

One of the disadvantages of the model presented in the previous subsection is that the infection state $z_{u,t+1}$ depends on all variables $z_{u,1},\ldots,z_{u,t}$ which does not factorize the model well. However, with a simple re-parameterization trick we can make the model completely Markovian in the sense that all variables at time $t+1$ only depend on variables at time $t$. In order to achieve this, let us introduce $|\mathcal{S}| \times T$ many latent variables $\mathcal{Y}:=\{y_{u,t}\}_{u\in\mathcal{S},t=1,\ldots,T}\in\{S,E_1,\ldots,E_M,I_1,\ldots,I_N,R\}^{|\mathcal{S}| \times T}$ where we assume that the maximum number of time steps that $z_{u,t}$ is in state $E$ equals $M$, and the maximum number of time steps that $z_{u,t}$ is in state $I$ is $N$. The meaning of the variable $y_{u,t}$ is then as follows
\begin{itemize}
    \item $y_{u,t}=S$: individual $u$ has not been infected and is susceptible,
    \item $y_{u,t}=E_m$: individual $u$ is infected but not contagious since $m$ time steps,
    \item $y_{u,t}=I_n$: individual $u$ is infected and is contagious since $n$ time steps,
    \item $y_{u,t}=R$: individual $u$ has recovered and is not contagious. 
\end{itemize}
Let us use the notation $\mathcal{Y}_{t}:=\{y_{u,t}\}_{u\in\mathcal{S}}$ to denote the set of latent states $y_{u,t}$ of all individuals $u \in \mathcal{S}$ at time $t$. Then, the only non-zero probabilities $P(y_{u,t+1}|\mathcal{Y}_t)$ are the transitions $S \rightarrow S$, $S \rightarrow E_1$, $E_m \rightarrow E_{m+1}$, $E_m \rightarrow I_1$, $I_n \rightarrow I_{n+1}$, $I_n \rightarrow R$, and $R \rightarrow R$ (see Figure~\ref{fig:markovian_graphical_model} for an example model): 
\begin{equation}
  P(y_{u,t+1}|\mathcal{Y}_t)  = \begin{cases}
    f(u,t,\mathcal{Y}_t) & \mbox{if } y_{u,t}=S \wedge y_{u,t+1}=S\\
    1-f(u,t,\mathcal{Y}_t)  & \mbox{if } y_{u,t}=S \wedge y_{u,t+1}=E_1\\
    1-\pi(m;q_E)  & \mbox{if } y_{u,t}=E_m \wedge y_{u,t+1}=E_{m+1}\\
    \pi(m;q_E)  & \mbox{if } y_{u,t}=E_m \wedge y_{u,t+1}=I_1\\
    1-\pi(n;q_I)  & \mbox{if } y_{u,t}=I_n \wedge y_{u,t+1}=I_{n+1}\\
    \pi(n;q_I)  & \mbox{if } y_{u,t}=I_n \wedge y_{u,t+1}=R\\
    1  & \mbox{if } y_{u,t}=R \wedge y_{u,t+1}=R\\
    0  & \mbox{otherwise}
    \end{cases}\,. \label{eq:conditional_model_y}
\end{equation}
Note that the function $f$ is identical to (\ref{eq:f_function}) except that the condition $z_{v,t}=I$ is replaced by $y_{v,t}\in \{I_1,\ldots,I_N\}$:
\begin{equation}
  f(u,t,\mathcal{Y}_t) = (1-p_0) \cdot \displaystyle\prod_{(v,u,t,\mathbf{x}) \in \mathcal{D}_{\mathrm{contact}}: y_{v,t} \in \{I_1,\ldots,I_N\}} \prod_{j=1}^J (1-p_j)^{x_j} \,. \label{eq:f_function_for_y}
\end{equation}
Similarly, the test outcomes are defined as follows
\begin{equation}
    P\left(o|y_{u,t}\right) =  
    \begin{cases}
    \alpha & \textrm{if}\ y_{u,t} \in \{I_1,\ldots,I_N\} \wedge o=0 \\
    1-\alpha & \textrm{if}\ y_{u,t} \in \{I_1,\ldots,I_N\} \wedge o=1 \\
    1-\beta & \textrm{if}\ y_{u,t}\in\left\{ S,E_1,\ldots,E_M,R\right\} \wedge o=0 \\
    \beta & \textrm{if}\ y_{u,t}\in\left\{ S,E_1,\ldots,E_M,R\right\} \wedge o=1  
    \end{cases}\,. \label{eq:test_outcome_for_y}
\end{equation}
With these definitons, it is easy to verify that
\begin{align*}
 P(z_{u,t}=S) & = P(y_{u,t}=S)\,, \\
 P(z_{u,t}=E) & = \sum_{m=1}^M P(y_{u,t}=E_m)\,, \\
 P(z_{u,t}=I) & = \sum_{n=1}^N P(y_{u,t}=I_n)\,, \\
 P(z_{u,t}=R) & = P(y_{u,t}=R) \,.
 \end{align*}
 Thus, the parameterization in terms of $y_{u,t}$ is more fine-grained than the parameterization of the CRISP model in terms of $z_{u,t}$.

\section{Inference in the CRISP Model}
\label{sec:inference}

For inference in the aforementioned model we are interested in computing the infection risk score of every individual $u$ at every time step $t$ given the test outcomes $\mathcal{O}$ available as well as the hyper-parameters ${\boldsymbol \theta}:=(p_0,p_1,\ldots,p_J)$ which cannot be set by knowledge of the diseases: the $J$ parameters $p_j$ represent the probabilities of COVID-19 infection transmission through the contact channel $j$ and $p_0$ captures the probability that an infection occurs at any time-step from exogenous influences.

In order to estimate ${\boldsymbol \theta}$, we will maximize the log-likelihood of the data $\mathcal{O}$, that is\footnote{Without loss of generality, we present here the formulation in terms of the four-state model presented in Subsection \ref{subsec:four-state-CRISP-model}. For the Markovian model presented in Subsection \ref{subsec:Markovian} the variable $\mathcal{Z}_T$ gets replaced with $\mathcal{Y}_T$.}
 \begin{align}
 {\boldsymbol \theta}^* & = \mathrm{argmax}_{\boldsymbol \theta} \log\left(P(\mathcal{O}|{\boldsymbol \theta})\right) \\ 
& = \mathrm{argmax}_{\boldsymbol \theta} \log\left(\sum_{\mathcal{Z}_T} P(\mathcal{Z}_T | \mathcal{O},{\boldsymbol \theta}) \cdot P(\mathcal{O}|{\boldsymbol \theta})\right) \\
  & = \mathrm{argmax}_{\boldsymbol \theta} \log\left(\sum_{\mathcal{Z}_T} P(\mathcal{Z}_T, \mathcal{O}|{\boldsymbol \theta})\right) \label{eq:full_likelihood} \,,
 \end{align}
where the second decomposition explicitly contains the posterior $P\left(\mathcal{Z}_T|\mathcal{O}\right)$. However, this posterior is not analytically tractable and therefore we will need to approximate it, either by performing block-Gibbs sampling of an infection trace $\mathbf{z}_u := (z_{u,1},\ldots,z_{u,T})$ of individual $u$ keeping all other infection traces $\left\{ \mathbf{z}_{v:v\neq u}\right\}$ fixed (Subsection \ref{subsec:infection-risk-score-inference}, \ref{subsec:efficient-block-Gibbs-sampling} and \ref{subsec:federated-block-Gibbs-sampling}) or by loopy-belief propagation (Subsection \ref{subsec:factor-graph-and-message-passing}).  
 
\subsection{Infection Risk Score Inference}
\label{subsec:infection-risk-score-inference}
Since we assume that the total number of days of the model, $T$, is not large\footnote{As of today, the COVID-19 pandemic is active for 90 days.}, we will enumerate all possible sequences of infection traces $\mathbf{z}_u$ and compute the un-normalized probability of $P\left(\mathbf{z}_{u}|\left\{ \mathbf{z}_{v:v\neq u}\right\},\mathcal{O},{\boldsymbol \theta}\right)$ for all terms that depend on elements of the trace $\mathbf{z}_u$ in order to re-normalize and draw from this distribution. Also, as our model is an SEIR model, we know that each sample $\mathbf{z}_u$ can be uniquely represented by a triple $\omega=(t_0,d_E,d_I) \in \mathbb{N}\times\mathbb{N}^+\times\mathbb{N}^+$ of time steps with $t_0$ being time steps individual $u$ is in state $S$, $d_E$ being time steps in state $E$, $d_I$ being time steps in state $I$ and the remaining $T-t_0-d_E-d_I$ being time steps in state $R$. 

There are three groups of factors that (might) involve $\mathbf{z}_u$ in the (un-normalized) conditional probability distribution 
$P\left(\mathbf{z}_{u}|\left\{ \mathbf{z}_{v:v\neq u}\right\},\mathcal{O},{\boldsymbol \theta}\right)$: 

\begin{align}
    \underbrace{\prod_{t=1}^{T-1} P(z_{u,t+1}|\mathcal{Z}_t)}_{A(\mathbf{z}_u)} \cdot \prod_{v\not= u}\underbrace{\prod_{t=1}^{T-1} P(z_{v,t+1}|\mathcal{Z}_t)}_{B(\mathbf{z}_u)} \cdot \underbrace{\prod_{(t,o)\in\mathcal{T}_u} P(o|z_{u,t})}_{C(\mathbf{z}_u)} \,. \label{eq:likelihood_function}
\end{align}
The first set of factors, $A(\mathbf{z}_u)$, captures the temporal evolution of the infection state changes of $\mathbf{z}_u$ directly and can be reduced to three factors based on $\omega$ and all the contacts $v$ that could have infected individual $u$. The second set of factors, $B(\mathbf{z}_u)$, captures the factors where the infectiousness of $u$ might impact other individuals $v$. Finally, the third set of factors, $C(\mathbf{z}_u)$, captures the outcome of tests on individual $u$. 

\paragraph{Factors $A(\mathbf{z}_u)$} 
In order to derive a compact representation of $A(\mathbf{z}_u)$, we assume that it can be written in terms of 
\begin{align}
    A(\mathbf{z}_u) & = l_0(t_0) \cdot l_E(d_E) \cdot l_I(d_I) \cdot l_\mathrm{infected} \label{eq:A_function}
\end{align}
Since the infectious status, $z_{v,t}=I$ of other individuals $v$ that had contact with $u$ only affect $u$ in the susceptible state, we can derive $l_0(t_0)$ and $l_\mathrm{infected}$ from (\ref{eq:conditional_model}) by collecting the $f$ terms (see (\ref{eq:f_function})): 
\begin{align}
    & \prod_{t=1}^{t_0-1} f(u,t,\mathcal{Z}_t) \cdot (1-f(u,t_0,\mathcal{Z}_{t_0})) & \\
    & = \left(\prod_{t=1}^{t_0-1} p_{u,t} \right) \cdot (1-p_0)^{t_0-1} \cdot (1-(1-p_0)p_{u,t_0}) & \\
    & = \underbrace{\left(\prod_{t=1}^{t_0-1} p_{u,t} \right)\cdot \left( \frac{1-(1-p_0)p_{u,t_0}}{p_0} \right)}_{l_\mathrm{infected}} \cdot \underbrace{(1-p_0)^{t_0-1}p_0}_{l_0(t_0)} \,, \label{eq:t_likelihood}
\end{align} 
where $p_{u,t} := \prod_{(v,u,t,\mathbf{x}) \in \mathcal{D}_{\mathrm{contact}}: z_{v,t}=I} \prod_j (1-p_j)^{x_j}$. Note that $l_0(t_0)$ is the density function of the geometric distribution. Similarly, given (\ref{eq:conditional_model}) and (\ref{eq:g_function}) we can derive $l_E(d_E)$ as
\begin{align}
    l_E(d_E) & = \prod_{d=1}^{d_E-1} (1-g(u,t_0+d,\mathcal{Z}_{u,t_0+d})) \cdot g(u,t_0+d_E,\mathcal{Z}_{u,t_0+d_E})) \nonumber \\
    & = \prod_{d=1}^{d_E-1} \left( 1-\frac{q_E(d)}{1-\sum_{i=1}^{d-1}q_E(i)} \right) \cdot \frac{q_E(d_E)}{1-\sum_{i=1}^{d_E-1}q_E(i)} \\ 
    & = \prod_{d=1}^{d_E-1} \left( \frac{1-\sum_{i=1}^{d}q_E(i)}{1-\sum_{i=1}^{d-1}q_E(i)} \right) \cdot \frac{q_E(d_E)}{1-\sum_{i=1}^{d_E-1}q_E(i)} \\
    & = q_E(d_E) \,.
\end{align} 
A similar derivation shows that $l_I(d_I) = q_I(d_I)$ which proves that the computational complexity of computing $A(\mathbf{z}_u)$ has been reduced to one factor for each contact during the $S$ states of $u$ and three additional factors corresponding to the compact representation $\omega$ for $\mathbf{z}_u$. The number of factors do not directly scale up with $T$.

\paragraph{Factors $B(\mathbf{z}_u)$} 
In order to derive a compact representation of $B(\mathbf{z}_u)$, we note that only the cases of $z_{v,t}=S$ potentially contain the value of $z_{u,t}$ for $v\not= u$ (see the value range of the function $g$ and $h$ in (\ref{eq:conditional_model})). In fact, looking at (\ref{eq:f_function}) it becomes evident that it requires $z_{u,t}=I$. Thus, $B(\mathbf{z}_u)$ is defined by
\begin{align}
    \prod_{t=1}^{T} \prod_{v\in\mathcal{C}_S(u,t)} f(v,t,\mathcal{Z}_t) \prod_{v\in\mathcal{C}_E(u,t)} (1-f(v,t,\mathcal{Z}_t))\,, \label{eq:B_function}
\end{align}
where $\mathcal{C}_S(u,t)$ and $\mathcal{C}_E(u,t)$ are the individuals that $u$ met at time $t$ who were susceptible and have either stayed susceptible or got exposed, respectively: 
\begin{align*}
    \mathcal{C}_S(u,t) & :=\{ v: (u,v,t,\mathbf{x}) \in \mathcal{D}_{\mathrm{contact}} \wedge z_{v,t}=S \wedge z_{v,t+1}=S\} \,, \\
    \mathcal{C}_E(u,t) & :=\{ v: (u,v,t,\mathbf{x}) \in \mathcal{D}_{\mathrm{contact}} \wedge z_{v,t}=S \wedge z_{v,t+1}=E\} \,.
\end{align*}

\subsection{Efficient Block-Gibbs Sampling}
\label{subsec:efficient-block-Gibbs-sampling}
In this subsection, we describe how we can speed up the block-Gibbs sampling step for drawing a sample infection trace $\mathbf{z}_u$ for individual $u$.

\paragraph{Constant terms} A key observation is that the term $C(\mathbf{z}_u) = \prod_{(t,o)\in\mathcal{T}_u} P(o|z_{u,t})$ in (\ref{eq:likelihood_function})---corresponding to test outcomes for individual $u$---is a constant for each infection trace $\mathbf{z}_u$ irrespective of the values of other infection traces $\{\mathbf{z}_{v: v\neq u}\}$. Thus, $C(\mathbf{z}_u)$ can be pre-computed at the start of the block-Gibbs sampling algorithm for individual $u$ and then reused every time we evaluate the likelihood of an infection trace $\mathbf{z}_u$. Similarly, the terms $l_0(t_0) = (1-p_0)^{t_0-1} p_0$, $l_E(d_E) = q_E(d_E)$ and $l_I(d_I) = q_I(d_I)$ in $A(\mathbf{z}_u)$ in (\ref{eq:A_function}) are constant for each infection trace $\mathbf{z}_u$ irrespective of the infection traces of other individuals. Hence, these terms can also be pre-computed at the start of the block-Gibbs sampling algorithm for individual $u$. 

\paragraph{Contacts into $u$} The remaining term $l_{\mathrm{infected}}$ in $A(\mathbf{z}_u)$ (see (\ref{eq:A_function})) captures the contribution due to contacts into individual $u$ from other (infectious) individuals $v$ who are in state $z_{v,t}=I$ prior to $u$ herself getting infected at $t_0$. As a result, $l_{\mathrm{infected}}$ depends on the values of other infection traces $\{\mathbf{z}_{v: v\neq u}\}$ and needs to be recomputed during each block-Gibbs sampling step to draw sample $\mathbf{z}_u$. Let us define 
\begin{eqnarray}
    l_{\mathrm{infected}}(t) & := & p_{u,t} \,, \label{eq:l_infected} \\
    l^\prime_{\mathrm{infected}}(t) & := & \frac{1 - (1-p_0)p_{u,t}}{p_0} \label{eq:l_infected_prime} \,,
\end{eqnarray}
where $p_{u,t} := \prod_{(v,u,t,\mathbf{x}) \in \mathcal{D}_{\mathrm{contact}}: z_{v,t}=I} \prod_j (1-p_j)^{x_j}$ (see also (\ref{eq:t_likelihood})). Then, we have 
\begin{eqnarray}
    l_{\mathrm{infected}} & = & \prod_{t=1}^{t_0-1} l_{\mathrm{infected}}(t) \cdot l^\prime_{\mathrm{infected}}(t_0) \label{eq:fast_l_infected_compute} 
\end{eqnarray} 
Thus, at the start of each block-Gibbs sampling step, we pre-compute (\ref{eq:l_infected}) and (\ref{eq:l_infected_prime}) for each time step $t$, and then use (\ref{eq:fast_l_infected_compute})  to compute $l_{\mathrm{infected}}$ for a particular infection trace $\mathbf{z}_u$. Note that this involves only $t_0$ multiplications (or additions in the log-domain).

\paragraph{Contacts out from $u$} We next turn our attention to computing $B(\mathbf{z}_u)$ for each infection trace $\mathbf{z}_u$ that captures the contribution due to contacts out from $u$. We introduce two states $\mathcal{Z}_t^I$ and $\mathcal{Z}_t^{\neg I}$ which are identical to $\mathcal{Z}_t$ except for the value of infection state $z_{u,t}$ which is $I$ is $\mathcal{Z}_t^I$ and one of $\{S, E, R\}$ in $\mathcal{Z}_t^{\neg I}$. Now, let 
\begin{equation}
    B(\mathbf{z}_u,t,\mathcal{Z}_t) := \prod_{v\in\mathcal{C}_S(u,t)} f(v,t,\mathcal{Z}_t) \prod_{v\in\mathcal{C}_E(u,t)} (1-f(v,t,\mathcal{Z}_t)) \label{eq:B_zu_t}
\end{equation} 
be the inner terms in (\ref{eq:B_function}). Note that for all contacts $(u,v,t,\mathbf{x})\in \mathcal{D}_{\mathrm{contact}}$, the terms $B(\mathbf{z}_u,t,\mathcal{Z}_t^I)$ and $B(\mathbf{z}_u,t,\mathcal{Z}^{\neg I})$ differ only in the factor $\prod_j (1-p_j)^{x_j}$ that is in $f(v,t,\mathcal{Z}_t^I)$ but not in $f(v,t,\mathcal{Z}_t^{\neg I})$ since $u$ is infectious at this time $t$ in $\mathcal{Z}_t^I$ but not in $\mathcal{Z}_t^{\neg I}$. Also, note that values of $\mathbf{z}_{u,t^\prime}$ for $t^\prime < t$ do not affect $B(\mathbf{z}_u,t,\mathcal{Z}_t)$. We can then obtain $B(\mathbf{z}_u)$ for each infection trace $\mathbf{z}_u$ value as 
\begin{equation}
    B(\mathbf{z}_u) := \mathrm{Constant}\cdot \prod_{t = t_0 + d_E}^{t_0 + d_E + d_I} \frac{ B(\mathbf{z}_u,t,\mathcal{Z}^I_t)}{ B(\mathbf{z}_u,t,\mathcal{Z}^{\neg I}_t)} \label{eq:B_zu_fast} \,,
\end{equation} 
where $\mathrm{Constant}$ is the product of $B(\mathbf{z}_u,t,\mathcal{Z}_t^{\neg I})$ over all time steps $t$ and can be ignored due to normalization of the sampling distribution. Again, the ratio $B(\mathbf{z}_u,t,\mathcal{Z}_t^I)/B(\mathbf{z}_u,t,\mathcal{Z}_t^{\neg I})$ can be pre-computed for all time steps $t$ at the start of the block-Gibbs sampling step for $\mathbf{z}_u$, and then used to compute $B(\mathbf{z}_u)$ for each infection trace $\mathbf{z}_u$ as in (\ref{eq:B_zu_fast}). 

\paragraph{Putting it all together} Note that the quantities $l_{\mathrm{infected}}(t)$, $l^\prime_{\mathrm{infected}}(t)$ and $B(\mathbf{z}_u,t,Z_t^I)/B(\mathbf{z}_u,t,\mathcal{Z}_t^{\neg I})$ only depend on the infection status of individual $u$ at time $t$ because the infection traces $\mathbf{z}_v$ of all other individuals $v$ are fixed when we are drawing a block-Gibbs sample for $\mathbf{z}_u$. Thus, the (un-normalized) conditional probability for each $\mathbf{z}_u$ value is obtained by taking the product of $A(\mathbf{z}_u)$,  $B(\mathbf{z}_u)$ and  $C(\mathbf{z}_u)$, which in turn are computed efficiently as described above from pre-computed values of $l_0(t_0)$, $l_E(d_E)$, $l_I(d_I)$ and $C(\mathbf{z}_u)$ at the start of the algorithm, and $l_{\mathrm{infected}}(t)$, $l^\prime_{\mathrm{infected}}(t)$ and $B(\mathbf{z}_u,t,\mathcal{Z}_t^I)/B(\mathbf{z}_u,t,\mathcal{Z}_t^{\neg I})$ for all time steps $t$ at the start of the block-Gibbs sampling step.

\paragraph{Additional implementation optimizations} We use two additional ideas to accelerate the implementation of the block-Gibbs sampling algorithm:
\begin{itemize}
    \item We never materialize the infection trace $\mathbf{z}_u$ because it is uniquely described by the triple $\omega=(t_0,d_E,d_I)$; each value $z_{u,t}$ can be computed by no more than three comparisons of $t$ with $t_0$, $t_0+d_E$ and $t_0+d_E+d_I$. Thus, the whole state of the latent variable model is represented by $3\times|\mathcal{S}|$ integers.
    \item We carry out all computations of probabilities in the log-domain so all functions become sums and products instead of products and powers. 
\end{itemize}

\paragraph{Block Glibbs Sampling Algorithm} Algorithm \ref{alg:gibbs_sampling} is block-Gibbs sampling algorithm for sampling $\mathcal{Z}^i_T$ from our CRISP model. It cycles through (random) individuals $u$, sampling the vector of latent variables ${\bf z}_u$ from the conditional distribution $P({\bf z}_u| \{{\bf z}_{v:v\neq u} \}, \mathcal{O}, {\boldsymbol \theta})$ until convergence. We can use the samples $\mathcal{Z}_T^1,\ldots,\mathcal{Z}_T^m$ drawn by this algorithm to compute the infection risk score for an individual $u$ at time $t$ by taking the fraction of samples $\mathcal{Z}_T^i$ in which the latent infection state $z_{u,t}\in\{E,I\}$. 

\begin{algorithm}[t]
    \DontPrintSemicolon
    \tcc{Initialization}
    Initialize each $\mathbf{z}_u = S \cdot \mathbf{1}$ \;
    \tcc{Precomputations independent of contact data}
    \ForAll{$(t_0,d_E,d_I) \in \mathbb{N}^+ \times \mathbb{N}^+ \times \mathbb{N}^+$ such that $t_0+d_E+d_I \leq T$} {
        Pre-compute $l_0(t_0)$, $l_E(d_E)$ and $l_I(d_I)$\; 
        Construct the sequence $\mathbf{z}_u$ with $t_0$ states $S$, $d_E$ states $E$, $d_I$ states $I$ and $T-t_0-d_E-d_I$ states $R$ \;
        Pre-compute $C(\mathbf{z}_u)$ according to (\ref{eq:likelihood_function}) \;
    }
     \Repeat{convergence}{
     Pick a random index $u$ \;
     \tcc{Precomputations dependent on contact data}
     \ForAll{time steps $t$}{
        Pre-compute $l_{\mathrm{infected}}(t)$ using (\ref{eq:l_infected}) and $l^\prime_{\mathrm{infected}}(t)$ using (\ref{eq:l_infected_prime}) \; 
        Pre-compute ratio $B(\mathbf{z}_u,t,\mathcal{Z}_t^I)/B(\mathbf{z}_u,t,\mathcal{Z}_t^{\neg I})$ using (\ref{eq:B_zu_t}) \;
     }
    \ForAll{$(t_0,d_E,d_I) \in \mathbb{N}^+ \times \mathbb{N}^+ \times \mathbb{N}^+$ such that $t_0+d_E+d_I \leq T$} {
        \tcc{Infection trace specific computations}
        Construct the sequence $\mathbf{z}_u$ with $t_0$ states $S$, $d_E$ states $E$, $d_I$ states $I$ and $T-t_0-d_E-d_I$ states $R$ \;
        Compute $\log(A(\mathbf{z}_u)) = \log l_0(t_0) + \log l_E(d_E) + \log l_I(d_I) + \log(l_{\mathrm{infected}})$ using (\ref{eq:fast_l_infected_compute}) \;
        Compute $\log(B(\mathbf{z}_u))$ using (\ref{eq:B_zu_fast}) \;
        Set $l_{t_0,d_E,d_I} =   \log(A(\mathbf{z}_u)) + \log(B(\mathbf{z}_u)) + \log(C(\mathbf{z}_u))$ \;
    }
    \tcc{Block-Gibbs sampling step}
    Sample $(t^*_0,d^*_E,d^*_I)$ with probability $\propto \exp(l_{t^*_0,d^*_E,d^*_I} - \max_{t_0,d_E,d_I}(l_{t_0,d_E,d_I}))$ \;
    Set $\mathbf{z}_u$ with ($S$,$E$,$I$,$R$) states corresponding to $(t^*_0,d^*_E,d^*_I)$ \;
    \Return{$\mathcal{Z}^i = \mathcal{Z}$} \;
    }

    \caption{Block-Gibbs sampling algorithm for CRISP model}
    \label{alg:gibbs_sampling}
 \end{algorithm}

\subsection{Hyperparameter Inference}
In order to estimate the hyper-parameters ${\boldsymbol \theta}$ of the CRISP model, would like to find ${\boldsymbol \theta}^*$ that maximizes the log-likelihood log (\ref{eq:full_likelihood}). However, since this is intractable, we propose to use the Monte Carlo Expectation-Maximization (EM) algorithm \cite{Bishop2006}. We will use EM to refine ${\boldsymbol \theta}$ in successive iterations. Let ${\boldsymbol \theta}_{\mathrm{old}}$ be the value of ${\boldsymbol \theta}$ computed in the previous iteration. Then, in the E step of the current iteration, we will estimate the expected complete-data log-likelihood
\begin{equation}
\sum_{\mathcal{Z}_T} P(\mathcal{Z}_T|\mathcal{O}, {\boldsymbol \theta}_\mathrm{old}) \cdot \log\left(P(\mathcal{Z}_T, \mathcal{O}|{\boldsymbol \theta})\right) \,.
\end{equation} 
We will use the block-Gibbs sampling procedure described in Algorithm \ref{alg:gibbs_sampling} to approximate the posterior distribution $P(\mathcal{Z}_T|\mathcal{O}, {\boldsymbol \theta}_\mathrm{old})$ over the latent infection status of individuals $u$. If the samples drawn from the posterior $P(\mathcal{Z}_T|\mathcal{O},{\boldsymbol \theta}_{\mathrm{old}})$ are $\mathcal{Z}_T^1,\ldots,\mathcal{Z}_T^m$, then in the M step, we will compute ${\boldsymbol \theta}$ that maximizes the expected complete-data log-likelihood
\begin{align}
{\boldsymbol \theta}_{\mathrm{next}} & = \mathrm{argmax}_{\boldsymbol \theta}\sum_{i=1}^m \log\left(P\left(\mathcal{Z}^i_T, \mathcal{O}|{\boldsymbol \theta}\right)\right) \\
& = \mathrm{argmax}_{\boldsymbol \theta}\sum_{i=1}^m \sum_{t=1}^{T-1} \sum_u \log\left(P\left(z^i_{u,t+1}, \mathcal{O}|\mathcal{Z}^i_t, {\boldsymbol \theta}\right)\right) \,,
\end{align}  
where $z_{u,t+1}^i$ is the infection state for individual $u$ at time $t+1$ in sample $\mathcal{Z}_T^i$. If $t_0^i$ denotes the number of initial $S$ states in the sample infection trace $\mathbf{z}_u^i$, we note that by virtue of (\ref{eq:conditional_model}) only the first $t_0^i$ terms depend on ${\boldsymbol \theta}$ which reduces the above maximization term to only
\begin{equation*}
\sum_{i=1}^m \sum_u \sum_{t=1}^{t_0^i-1} \log\left(f\left(u^i,t,\mathcal{Z}^i_t|{\boldsymbol \theta}\right)\right) + \log\left(1-f\left(u^i,t_0^i,\mathcal{Z}^i_t|{\boldsymbol \theta}\right) \right) \,.
\end{equation*}  
We use stochastic gradient descent to compute the ${\boldsymbol \theta}$ values that maximize the above expression. We also note that for numerical stability, we re-parameterize $p_j$ via $w_j$ as $p_j=\exp(w_j)/(1+\exp(w_j))$ which allows for an unconstrained optimization over $\mathbf{w}$.

\subsection{Federated Block-Gibbs Sampling}
\label{subsec:federated-block-Gibbs-sampling}
We can extend the block-Gibbs sampling algorithm in CRISP to a federated learning setting \cite{McMahan2017} where local contact and test outcome data for an individual $u$ are utilized to compute the block-Gibbs sample $\mathbf{z}_u$ on the individual’s mobile device {\em without} ever needing to be shared with anyone else. This has two benefits: (1) We distribute the block-Gibbs sampling algorithm across hundreds of millions of mobile devices in the world and thereby utilize their distributed computational power, and (2) Contact and test outcome data for an individual are stored only on the individual’s mobile device and not shared with other mobile devices–--this preserves a user's privacy. In the federated setting, the contact data is never centralized---instead for each individual $u$, her device executes the block-Gibbs sampling step to draw sample $\mathbf{z}_u$ only using the locally available contacts and test outcome data for $u$, as well as additional  “minimal statistics” sent to $u$ by the devices of its past contacts. In the following two paragraphs, we explain how to compute the factors $A(\mathbf{z}_u)$, $B(\mathbf{z}_u)$ and $C(\mathbf{z}_u)$ in (\ref{eq:likelihood_function}) in a federated setting (see Algorithm~\ref{alg:federated_gibbs_sampling} for the pseudo-code which runs on every mobile device).

\paragraph{Factors $A(\mathbf{z}_u)$ and $C(\mathbf{z}_u)$}
A key observation is that the terms $l_0(t_0)$, $l_E(d_E)$, $l_I(d_I)$ in $A(\mathbf{z}_u)$ as well as the factor $C(\mathbf{z}_u)$ can all be computed locally on the device with the contact and test outcome information available on the device. In order to compute the remaining term $l_{\mathrm{infected}}$ in $A(\mathbf{z}_u)$, we only require information on the individuals $v$ who had a contact with $u$ at each time step $t$ and the infection status $z_{v,t}$ of $v$ at the time of the contact. Individual $u$’s mobile device already has the contact information for $u$; thus all that is required to compute $l_{\mathrm{infected}}$ are the current infection traces $\mathbf{z}_v$ for all individuals $v$ who have had contacts with $u$. Since each infection trace is uniquely characterized by a $(t_0,d_I,d_E)$ triple, we require the mobile devices of all individuals $v$ who have had a contact with $u$ to send $u$’s device the $(t_0,d_I,d_E)$ triple corresponding to $\mathbf{z}_v$.

\paragraph{Factor $B(\mathbf{z}_u)$}
In order to compute $B(\mathbf{z}_u)$ as defined in (\ref{eq:B_function}), we require the term $f(v,t,\mathcal{Z}_t)$ for each individual $v$ who has had a contact with $u$ at time $t$ and whose infection state $z_{v,t} = S$. Let $f_{-u}(v,t,\mathcal{Z}_t)$ be defined as in (\ref{eq:f_function}) over all contacts of $v$ at time $t$ except for individual $u$. Then, the device for each individual $v$ who has had a contact with $u$ at time $t$ and whose infection state $z_{v,t} = S$ sends to $u$’s device the quantity $f_{-u}(v,t,\mathcal{Z}_t)$ computed based on $v$’s view of the infection traces of its contacts. These terms are used by $u$’s device to compute $B(\mathbf{z}_u)$ as defined in (\ref{eq:B_function}).

\begin{algorithm}[t]
    \DontPrintSemicolon
    \tcc{Initialization}
    Initialize $\mathbf{z}_u = S \cdot \mathbf{1}$ \;
    Initialize $\mathcal{N}_v = (\infty,0,0,\emptyset)$ for all $v$ \tcp{Stores minimal statistics from contacts}\; 
    \tcc{Precomputations independent of contact data}
    \Repeat{forever}{
        \tcc{Update minimal statistic received in the incoming queue}
        \ForAll{$\{(t^v_0,d^v_E,d^v_I,\{f_{-u}(v,t,\mathcal{Z}_t)\})\}_v$ in the incoming message queue} {
            $\mathcal{N}_v \leftarrow (t^v_0,d^v_E,d^v_I,\{f_{-u}(v,t,\mathcal{Z}_t)\})$ \; 
        }
        \tcc{Precomputations of test outcomes}
        \ForAll{$(t_0,d_E,d_I) \in \mathbb{N}^+ \times \mathbb{N}^+ \times \mathbb{N}^+$ such that $t_0+d_E+d_I \leq T$} {
            Pre-compute $l_0(t_0)$, $l_E(d_E)$ and $l_I(d_I)$\; 
            Pre-compute $C(\mathbf{z}_u)$ for this sequence according to (\ref{eq:likelihood_function}) \;
        }
        \tcc{Precomputations dependent on contact data}
        \ForAll{time steps $t$}{
            Pre-compute $l_{\mathrm{infected}}(t)$ using (\ref{eq:l_infected}) and $l^\prime_{\mathrm{infected}}(t)$ using (\ref{eq:l_infected_prime}) and $(t^v_0,d^v_E,d^v_I)$ in $\mathcal{N}_v$ for all past contacts $v$\; 
            Pre-compute ratio $B(\mathbf{z}_u,t,\mathcal{Z}_t^I)/B(\mathbf{z}_u,t,\mathcal{Z}_t^{\neg I})$ using (\ref{eq:B_zu_t}) and $\{f_{-u}(v,t,\mathcal{Z}_t)\}$ in $\mathcal{N}_v$ for all past contacts $v$\;
        }
        \ForAll{$(t_0,d_E,d_I) \in \mathbb{N}^+ \times \mathbb{N}^+ \times \mathbb{N}^+$ such that $t_0+d_E+d_I \leq T$} {
            \tcc{Infection trace specific computations}
            Construct the sequence $\mathbf{z}_u$ with $t_0$ states $S$, $d_E$ states $E$, $d_I$ states $I$ and $T-t_0-d_E-d_I$ states $R$ \;
            Compute $\log(A(\mathbf{z}_u)) = \log l_0(t_0) + \log l_E(d_E) + \log l_I(d_I) + \log(l_{\mathrm{infected}})$ using (\ref{eq:fast_l_infected_compute}) \;
            Compute $\log(B(\mathbf{z}_u))$ using (\ref{eq:B_zu_fast}) \;
            Set $l_{t_0,d_E,d_I} = \log(A(\mathbf{z}_u)) + \log(B(\mathbf{z}_u)) + \log(C(\mathbf{z}_u))$ \;
        }
        \tcc{Block-Gibbs sampling step}
        Sample $(t^*_0,d^*_E,d^*_I)$ with probability $\propto \exp(l_{t^*_0,d^*_E,d^*_I} - \max_{t_0,d_E,d_I}(l_{t_0,d_E,d_I}))$ \;
        Set $\mathbf{z}_u$ with ($S$,$E$,$I$,$R$) states corresponding to $(t^*_0,d^*_E,d^*_I)$ \;
        \tcc{Send minimal statistic to all contacts}
        \ForAll{$v$ in past contact list} {
            $\mathcal{F} = \{f_{-v}(u,t,\mathcal{Z}_t) : (u,v,t,x) \in \mathcal{D}_{\mathrm{contact}} \wedge t \leq t_0^*\}$\;
            Send message $(t^*_0,d^*_E,d^*_I,\mathcal{F})$ to $v$\; 
        }
    }
    \caption{Federated block-Gibbs Sampling algorithm for CRISP model}
    \label{alg:federated_gibbs_sampling}
 \end{algorithm}

 \subsection{Loopy Belief Propagation}
 \label{subsec:factor-graph-and-message-passing}

In this section, we derive an efficient inference algorithm for the Markovian CRISP model introduced in Subsection{subsec:Markovian} using {\em factor graphs} \cite{KscFreLoe2001}. The factor graph for the posterior distribution has two types of nodes: (1) nodes representing the {\it variables} which are the latent infection states $y_{u,t}$ and the test outcomes $o_{u,t}$, and (2) nodes representing the {\it  factors}, $f_{u,t}$ corresponding to $P(y_{u,t+1}|\mathcal{Y}_t)$ as defined in (\ref{eq:conditional_model_y}) and $g_{u,t}$ corresponding to $P(o|y_{u,t})$ as defined in (\ref{eq:test_outcome_for_y}). The factor graph contains undirected edges connecting each factor node to all of the variable nodes on which that factor depends. Thus, each factor $f_{u,t}$ is a function of variables $\mathbf{y}^f_{u,t} = \{y_{u,t}, y_{u,t+1}\} \cup \{y_{v,t}: (v, u, t,\mathbf{x})\in \mathcal{D}_{\mathrm{contact}}\}$, and each factor $g_{u,t}$ is a function of variables $\mathbf{y}^g_{u,t} = \{o_{u,t}, y_{u,t}\}$. Finally, for a node $n$ in the factor graph, we will denote by $N(n)$ the neighbors of $n$. 

Using this notation, the posterior $P(\mathcal{Y}_t|\mathcal{O})$ is given by
\begin{eqnarray}
P(\mathcal{Y}_T|\mathcal{O}) & \propto & \prod_t \prod_u f_{u, t}(\mathbf{y}^f_{u,t}) \cdot \prod_{(u, t, o)\in \mathcal{D}_{\mathrm{test}}} g_{u,t}(\mathbf{y}^g_{u,t})\,.
\end{eqnarray}
We are interested in computing the marginal $P(y_{u,t}|\mathcal{O}) = \sum_{\mathcal{Y}_T\backslash \{y_{u,t}\}} P(\mathcal{Y}_T|\mathcal{O})$, where $\mathcal{Y}_T\backslash \{y_{u,t}\}$ denotes the set of variables in $\mathcal{Y}_T$ with $y_{u,t}$ omitted. In belief propagation, the marginal of any variable is computed by propagating messages between variable and factor nodes. In order to evaluate the message sent by a variable node $n\in \mathbf{y}^f_{u,t}$ to an adjacent factor node $f_{u,t}$ along the connecting link, we simply take the product of the incoming messages along all of the other links.

\begin{eqnarray}
\mu_{n\rightarrow f_{u,t}}(n) = \prod_{n^\prime\in N(n)\backslash \left\{f_{u,t}\right\}} \mu_{n^\prime\rightarrow n}(n) \,. \label{eq:message-variable-to-factor}
\end{eqnarray}

In order to compute the message sent by a factor node $f_{u,t}$ to a variable node $n\in \mathbf{y}^f_{u, t}$, we take the product of the incoming messages from the other variables connected to this factor node and multiply by the factor function associated with that node before finally marginalizing over all of the variables associated with the factor.

\begin{eqnarray}
\mu_{f_{u,t}\rightarrow n}(n) = \sum_{\mathbf{y}^f_{u,t}\backslash\left\{ n \right\}} f_{u,t}(\mathbf{y}^f_{u,t}) \cdot \prod_{n^\prime\in N(f_{u,t})\backslash\left\{n\right\}} \mu_{n^\prime \rightarrow f_{u,t}}(n') \,. \label{eq:message-factor-to-variable}
\end{eqnarray}

Note that every node $o_{u, t}$ in the factor graph is a leaf, and so the message $\mu_{o_{u,t}\rightarrow g_{u,t}} = 1$. Thus, we get that the message from factor node $g_{u, t}$ to variable node $y_{u, t}$ is given by: 
\begin{eqnarray}
\mu_{g_{u, t}\rightarrow y_{u,t}}(y_{u,t}) = g_{u,t}(\mathbf{y}^g_{u,t}) = P(o_{u,t}|y_{u,t})\,. \label{eq:message-gut-yut}
\end{eqnarray}

Since the factor graph contains loops, we need to define a message passing schedule. To resolve loops, we pass an initial message given by the unit function across every link in each direction. Each node transmits a message along a link once it has received incoming messages along all of the other links. Once the algorithm has converged, or has run for a specified amount of time, the marginal for variable node $y_{u,t}$  is computed using the product of the most recently received incoming messages along all of the links from neighboring factor nodes $n\in N(y_{u, t})$  into node $y_{u, t}$. Thus, we get
\begin{eqnarray}
P(y_{u, t}|\mathcal{O}) = \prod_{n\in N(y_{u,t})} \mu_{n\rightarrow y_{u,t}}(y_{u,t})\,.
\end{eqnarray}

\begin{figure}
    \centering
    \begin{tikzpicture}[scale=1.2,node distance=1.0 cm and 1.3 cm]
        \node[factor, label=below:\tiny $f_{1,0}$] (f10) at (0,0) {};
        \node[var,thick]    (y11) [right=of f10] {$y_{1,1}$}; 
        \node[factor, label=below:\tiny $f_{1,1}$] (f11) [right=of y11] {};
        \node[var,thick]    (y12) [right=of f11] {$y_{1,2}$};  
        \node[factor, label=below:\tiny $f_{1,2}$] (f12) [right=of y12] {};
        \node[var,thick]    (y13) [right=of f12] {$y_{1,3}$};  
    
        \node[factor, label=below:\tiny $f_{2,0}$] (f20) at (0,-2.5)    {};
        \node[var,thick]    (y21) [right=of f20] {$y_{2,1}$}; 
        \node[factor, label=below:\tiny $f_{2,1}$] (f21) [right=of y21] {};
        \node[var,thick]    (y22) [right=of f21] {$y_{2,2}$};  
        \node[factor, label=below:\tiny $f_{2,2}$] (f22) [right=of y22] {};
        \node[var,thick]    (y23) [right=of f22] {$y_{2,3}$};  
    
        \node[factor, label=below:\tiny $f_{3,0}$] (f30) at (0,-5)      {};
        \node[var,thick]    (y31) [right=of f30] {$y_{3,1}$}; 
        \node[factor, label=below:\tiny $f_{3,1}$] (f31) [right=of y31] {};
        \node[var,thick]    (y32) [right=of f31] {$y_{3,2}$};  
        \node[factor, label=below:\tiny $f_{3,2}$] (f32) [right=of y32] {};
        \node[var,thick]    (y33) [right=of f32] {$y_{3,3}$};  
    
        \node[factor, label=right:\tiny $g_{1,2}$] (g12) [above=of y12] {};
        \node[var,thick,fill=gray] (o12) [above=of g12] {$o_{1,2}$};  
    
        \path[-] (f10) edge[thick] (y11);
        \path[-] (f11) edge[thick] (y11);
        \path[-] (f11) edge[thick] (y12);
        \path[-] (f12) edge[thick] (y12);
        \path[-] (f12) edge[thick] (y13);
    
        \path[-] (f20) edge[thick] (y21);
        \path[-] (f21) edge[thick] (y21);
        \path[-] (f21) edge[thick] (y22);
        \path[-] (f22) edge[thick] (y22);
        \path[-] (f22) edge[thick] (y23);
    
        \path[-] (f30) edge[thick] (y31);
        \path[-] (f31) edge[thick] (y31);
        \path[-] (f31) edge[thick] (y32);
        \path[-] (f32) edge[thick] (y32);
        \path[-] (f32) edge[thick] (y33);
    
        \path[-] (f12) edge[thick] (y22);
        \path[-] (f22) edge[thick] (y12);
        \path[-] (f22) edge[thick] (y32);
        \path[-] (f32) edge[thick] (y22);
        \path[-] (g12) edge[thick] (y12);
        \path[-] (g12) edge[thick] (o12);
    
        \draw[black] (f10) to[bend left=45] node[pos=0.55] {\tiny $\mu_{f_{1,0}\rightarrow y_{1,1}}$}  (y11);
        \draw[red] (f11) to[bend left=45] node[pos=0.55] {\tiny $\mu_{f_{1,1}\rightarrow y_{1,1}}$}  (y11);
        \draw[blue] (f11) to[bend left=45] node[pos=0.55] {\tiny $\mu_{f_{1,1}\rightarrow y_{1,2}}$}  (y12);
        \draw[red] (f12) to[bend left=45] node[pos=0.55] {\tiny $\mu_{f_{1,2}\rightarrow y_{1,2}}$}  (y12);
        \draw[blue] (f12) to[bend left=45] node[pos=0.55] {\tiny $\mu_{f_{1,2}\rightarrow y_{1,3}}$}  (y13);
    
        \draw[black] (f20) to[bend left=45] node[pos=0.55] {\tiny $\mu_{f_{2,0}\rightarrow y_{2,1}}$}  (y21);
        \draw[red] (f21) to[bend left=45] node[pos=0.55] {\tiny $\mu_{f_{2,1}\rightarrow y_{2,1}}$}  (y21);
        \draw[blue] (f21) to[bend left=45] node[pos=0.55] {\tiny $\mu_{f_{2,1}\rightarrow y_{2,2}}$}  (y22);
        \draw[red] (f22) to[bend left=45] node[pos=0.55] {\tiny $\mu_{f_{2,2}\rightarrow y_{2,2}}$}  (y22);
        \draw[blue] (f22) to[bend left=45] node[pos=0.55] {\tiny $\mu_{f_{2,2}\rightarrow y_{2,3}}$}  (y23);
    
        \draw[black] (f30) to[bend left=45] node[pos=0.55] {\tiny $\mu_{f_{3,0}\rightarrow y_{3,1}}$}  (y31);
        \draw[red] (f31) to[bend left=45] node[pos=0.55] {\tiny $\mu_{f_{3,1}\rightarrow y_{3,1}}$}  (y31);
        \draw[blue] (f31) to[bend left=45] node[pos=0.55] {\tiny $\mu_{f_{3,1}\rightarrow y_{3,2}}$}  (y32);
        \draw[red] (f32) to[bend left=45] node[pos=0.55] {\tiny $\mu_{f_{3,2}\rightarrow y_{3,2}}$}  (y32);
        \draw[blue] (f32) to[bend left=45] node[pos=0.55] {\tiny $\mu_{f_{3,2}\rightarrow y_{3,3}}$}  (y33);
    
        \draw[brown] (f12) to[bend left=45] node[pos=0.45,sloped] {\tiny $\mu_{f_{1,2}\rightarrow y_{2,2}}$}  (y22);
        \draw[brown] (f32) to[bend left=45] node[pos=0.45,sloped] {\tiny $\mu_{f_{3,2}\rightarrow y_{2,2}}$}  (y22);
        \draw[brown] (f22) to[bend left=45] node[pos=0.45,sloped] {\tiny $\mu_{f_{2,2}\rightarrow y_{1,2}}$}  (y12);
        \draw[brown] (f22) to[bend left=45] node[pos=0.45,sloped] {\tiny $\mu_{f_{2,2}\rightarrow y_{3,2}}$}  (y32);
    
        \draw[teal] (g12) to[bend left=45] node[pos=0.5] {\tiny $\mu_{g_{1,2}\rightarrow y_{1,2}}$}  (y12);
    
    
    \end{tikzpicture} 
    \caption{Factor graph of the Markovian CRISP contact infection spread model for 3 people over 3 time steps where individual $u=2$ meets both individual $u=1$ and $u=2$ at time $t=2$ and one test outcome of individual $u=1$ at time $t=2$. Note that we also show all factor-to-variable messages denoting the different types of messages in black, blue, red, brown and teal. Note also that the factor $f_{u,t}$ is given by $P(y_{u,t+1}|\mathcal{Y}_t)$ in (\ref{eq:conditional_model_y}) and the factor $g_{u,t}$ is given by $P\left(o_{u,t}|y_{u,t}\right)$ in (\ref{eq:test_outcome_for_y}). \label{fig:factor_graph}}
\end{figure}

An example factor graph together with all messages from factors to variables is shown in Figure~\ref{fig:factor_graph}. In the following three subsections, we will derive efficient message equations for all messages emitting from the factor $f_{u,t}$, namely\footnote{In order to simplify notation, let the individuals with whom $u$ has had contact at time $t$ be $v_1,v_2,\ldots,v_r$ and the corresponding feature vectors be $\mathbf{x}_1,\ldots,\mathbf{x}_r$. Thus, $\mathbf{y}^f_{u,t} = \{y_{u,t+1},y_{u,t},y_{v_1,t},\ldots,y_{v_r,t}\}$.}:
\begin{itemize}
    \item $\mu_{f_{u,t}\rightarrow y_{u,t+1}}$ (blue arrows): Message from past infection states $y_{u,t},y_{v_1,t},\ldots,y_{v_r,t}$ to future infection state $y_{u,t+1}$.
    \item $\mu_{f_{u,t}\rightarrow y_{u,t}}$ (red arrows): Message from future infection state $y_{u,t+1}$ and contact infection states $y_{v_1,t},\ldots,y_{v_r,t}$ to past infection state $y_{u,t}$, and 
    \item $\mu_{f_{u,t}\rightarrow y_{v_r,t}}$ (brown arrows): Message from past infection state $y_{u,t}$, future infection state $y_{u,t+1}$ and other contact infection states $y_{v_1,t},\ldots,y_{v_{r-1},t}$ to contact infection state $y_{v_r,t}$
\end{itemize}
The message $\mu_{g_{u, t}\rightarrow y_{u,t}}$ is already given in (\ref{eq:message-gut-yut}) (teal arrows) and the message $\mu_{f_{u,0}\rightarrow y_{u,1}}$ is simply the prior probability for infection state of each individual (black arrows).

\paragraph{Reduction from $O((M+N+2)^r)$ to $O(2^r)$} We note that the message equation $\mu_{f_{u,t}\rightarrow n}$ in (\ref{eq:message-factor-to-variable}) requires marginalizing over all of the $r+1$ variables in $\mathbf{y}^f_{u,t}\backslash\left\{n\right\}$. Since each latent variable $y_{u,t}$ can be in any one of $M+N+2$ infection states in $\{S, E_1,\ldots, E_M, I_1, \ldots, I_N, R\}$, the time complexity to compute the message in a naive way is $(M+N+2)^{r+1}$. However, looking closely at the functional form of factor $f_{u,t}$ (see (\ref{eq:conditional_model_y})), we notice that it only depends on all $r$ variables $y_{v_i,t}$ whenever $y_{u,t}=S$ and $y_{u,t+1} \in \{S,E_1\}$. In fact, in this case the function $f(u,t,\mathcal{Y}_t)$ as defined in (\ref{eq:conditional_model_y}) takes the same values for $y_{v_i,t} \in \mathcal{I}$ and $y_{v_i,t} \in \mathcal{N}$ where $\mathcal{I} := \{I_1, \ldots, I_N\}$ and $\mathcal{N} := \{S,E_1, \ldots, E_M,R\}$. Hence, we can apply the following simplification
\begin{eqnarray}
    & & \sum_{y_{v_1,t}}\cdots \sum_{y_{v_r,t}} f_{u,t}(S,y_{u,t+1},y_{v_1,t},\ldots,y_{v_r,t}) \cdot \prod_{i=1}^r \mu_{y_{v_i,t} \rightarrow f_{u,t}}(y_{v_i,t}) \\
    & = & \sum_{Y_1 \in \{\mathcal{I},\mathcal{N}\}} \cdots \sum_{Y_r \in \{\mathcal{I},\mathcal{N}\}} f_{u,t}(S,y_{u,t+1},\psi(Y_1),\ldots,\psi(Y_r)) \cdot \prod_{i=1}^r \mu_{y_{v_i,t} \rightarrow f_{u,t}}(Y_i) \,,
\end{eqnarray}
where we used 
\begin{eqnarray}
    \mu_{y_{v_i,t}\rightarrow f_{u,t}} (\mathcal{I}) & := & \sum_{n=1}^N \mu_{y_{v_i,t}\rightarrow f_{u,t}}(I_n)\,, \label{eq:coalesced_message_I} \\
    \mu_{y_{v_i,t}\rightarrow f_{u,t}} (\mathcal{N}) & := & \mu_{y_{v_i,t}\rightarrow f_{u,t}}(S) + \mu_{y_{v_i,t}\rightarrow f_{u,t}}(R) + \sum_{m=1}^M \mu_{y_{v_i,t}\rightarrow f_{u,t}}(E_m) \label{eq:coalesced_message_N} \,, \\
    \psi(q) & := & \begin{cases}
        I_1 & \mbox{if } q=\mathcal{I} \\
        S  & \mbox{if } q=\mathcal{N}
        \end{cases} \,.
\end{eqnarray} 
Note that this simplification reduces the computational complexity of the message equation from $(M+N+2)^r$ to $2^r$ summations.

\paragraph{Reduction from $O(2^r)$ to $O(r)$} In order to reduce the computational complexity further, note that the function $f_{u,t} = P(y_{u,t+1}|\mathcal{Y}_t)$ as defined in (\ref{eq:conditional_model_y}) is the sum of functions that have a particular factorizing structure 
\begin{equation}
    f_{u,t}(y_{u,t},y_{u,t+1},y_{v_1,t},\ldots,y_{v_r,t}) = \sum_k \kappa_k(y_{u,t},y_{u,t+1}) \cdot \prod_{i=1}^r a_{i,k}^{\mathbb{I}\left(y_{v_i,t} \in \mathcal{I}\right)} \cdot b_{i,k}^{\mathbb{I}\left(y_{v_i,t} \in \mathcal{N}\right)} \,.
\end{equation}
In Lemma~\ref{lem:expansion} in the appendix, we show that in this case the summation over the $2^r$ values of the $Y_i \in \{\mathcal{I},\mathcal{N}\}$ reduces to a simple product such that  
\begin{eqnarray}
    & &\hspace*{-2cm} \sum_{Y_1 \in \{\mathcal{I},\mathcal{N}\}} \cdots \sum_{Y_r \in \{\mathcal{I},\mathcal{N}\}} \kappa(y_{u,t},y_{u,t+1}) \cdot \prod_{i=1}^r a_i^{\mathcal{Y}_i = \mathcal{I}} \cdot b_i^{\mathcal{Y}_i = \mathcal{N}} \cdot \mu_{y_{v_i,t} \rightarrow f_{u,t}}(Y_i) \\
    & = & \kappa(y_{u,t},y_{u,t+1}) \cdot \prod_{i=1}^r \left(a_i \cdot \mu_{y_{v_i,t} \rightarrow f_{u,t}}(\mathcal{I}) + b_i \cdot \mu_{y_{v_i,t} \rightarrow f_{u,t}}(\mathcal{N}) \right) \label{eq:linear_reduction_template} \,.
\end{eqnarray}
\paragraph{Special Cases} We often encounter the two special cases of $\kappa = a_i = b_i = 1$, and $\kappa = 1-p_0$, $a_i = \prod_{j=1}^J (1-p_j)^{x_{ij}}$, and $b_i = 1$, respectively. Using (\ref{eq:linear_reduction_template}), they result in these frequently occurring constant for the marginalization over all $(M+N+2)$ states of the $r$ variables $y_{v_1,t},\ldots,y_{v_r,t}$
\begin{eqnarray}
    A & := & \prod_{i=1}^r \left(\mu_{y_{v_i,t}\rightarrow f_{u,t}}(\mathcal{I}) + \mu_{y_{v_i,t}\rightarrow f_{u,t}}(\mathcal{N})\right) \label{eq:A} \,, \\
    B & := & (1-p_0) \cdot \prod_{i=1}^r \left(\mu_{y_{v_i,t}\rightarrow f_{u,t}}(\mathcal{I}) \cdot \prod_{j=1}^J (1-p_j)^{x_{ij}} + \mu_{y_{v_i,t}\rightarrow f_{u,t}}(\mathcal{N}) \right) \label{eq:B} \,.
\end{eqnarray}

\subsubsection{Message $\mu_{f_{u,t}\rightarrow y_{u,t+1}}$} 

\paragraph{Case $y_{u,t+1} = S$} In this case, $f_{u,t}(\mathbf{y}^f_{u,t})$ is non-zero with value $f(u,t,\mathcal{Y}_t)$ only for $y_{u,t} = S$. Thus, using (\ref{eq:linear_reduction_template}) and (\ref{eq:B}) we get
\begin{equation}
    \mu_{f_{u,t}\rightarrow y_{u,t+1}}(S) = \mu_{y_{u,t}\rightarrow f_{u,t}}(S) \cdot B \label{eq:fut-yutp1-case-s-equation}
\end{equation}

\paragraph{Case $y_{u, t+1} = E_1$} In this case, $f_{u,t}(\mathbf{y}^f_{u,t})$ is non-zero with value $1-f(u,t,\mathcal{Y}_t)$ only for $y_{u,t} = S$. Thus, using (\ref{eq:linear_reduction_template}), (\ref{eq:A}) and (\ref{eq:B}) we get  
\begin{equation}
    \mu_{f_{u,t}\rightarrow y_{u,t+1}}(E_1) = \mu_{y_{u,t}\rightarrow f_{u,t}}(S) \cdot (A - B)\,.
\end{equation}

\paragraph{Case $y_{u,t+1} = E_{m+1}, 1\leq m\leq M-1$} In this case, $f_{u,t}(\mathbf{y}^f_{u,t})$ is non-zero with value $1-\pi(m;q_E)$ only for $y_{u,t} = E_m$ and independent of the infection states $y_{v_i,t}$ of contacts $v_i$ at time $t$. Thus, using (\ref{eq:A}) we get 
\begin{equation}
\mu_{f_{u,t}\rightarrow y_{u,t+1}}(E_{m+1}) = \mu_{y_{u,t}\rightarrow f_{u,t}}(E_{m}) \cdot (1-\pi(m; q_E))\cdot A  
\end{equation}

\paragraph{Case $y_{u,t+1} = I_1$} In this case, $f_{u,t}(\mathbf{y}^f_{u,t})$ is non-zero with value $\pi(m;q_E)$ only for $y_{u,t} = E_m$, $1\leq m \leq M$ and independent of the infection states $y_{v_i,t}$ of contacts $v_i$ at time $t$. Thus, using (\ref{eq:A}) we get  
\begin{equation}
\mu_{f_{u,t}\rightarrow y_{u,t+1}}(I_1) = \left(\sum_{m=1}^M \mu_{y_{u,t}\rightarrow f_{u,t}}(E_{m}) \cdot \pi(m; q_E) \right) \cdot A
\end{equation}

\paragraph{Case $y_{u,t+1} = I_{n+1}, 1\leq n\leq N-1$} In this case, $f_{u,t}(\mathbf{y}^f_{u,t})$ is non-zero with value $1-\pi(n;q_I)$ only for $y_{u,t} = I_n$ and independent of the infection states $y_{v_i,t}$ of contacts $v_i$ at time $t$. Thus, using (\ref{eq:A}) we get 
\begin{equation}
\mu_{f_{u,t}\rightarrow y_{u,t+1}}(I_{n+1}) = \mu_{y_{u,t}\rightarrow f_{u,t}}(I_{n}) \cdot (1-\pi(n; q_I)) \cdot A
\end{equation}

\paragraph{Case $y_{u,t+1} = R$} In this case, $f_{u,t}(\mathbf{y}^f_{u,t})$ is non-zero with value $\pi(n;q_I)$ only for $y_{u,t} = I_n$, $1\leq n \leq N$ or with value $1$ for $y_{u,t}=R$. Also, $f_{u,t}(\mathbf{y}^f_{u,t})$ is independent of the infection states $y_{v_i,t}$ of contacts $v_i$ at time $t$. Thus, using (\ref{eq:A}) we get 
\begin{equation}
\mu_{f_{u,t}\rightarrow y_{u,t+1}}(R) = \left(\mu_{y_{u,t}\rightarrow f_{u,t}}(R) + \sum_{n=1}^N \pi(n;q_I)\cdot \mu_{y_{u,t}\rightarrow f_{u,t}}(I_{n})\right) \cdot A
\end{equation}

\subsubsection{Message $\mu_{f_{u,t}\rightarrow y_{u,t}}$} 

\paragraph{Case $y_{u,t} = S$} In this case, $f_{u,t}(\mathbf{y}^f_{u,t})$ is non-zero with value $f(u,t,\mathcal{Y}_t)$ for $y_{u,t+1} = S$, and with value $1-f(u,t,\mathcal{Y}_t)$ for $y_{u,t+1} = E_1$, respectively. Thus, using (\ref{eq:linear_reduction_template}) repeatedly as well as (\ref{eq:A}) and (\ref{eq:B}), we get 
\begin{equation}
    \mu_{f_{u,t}\rightarrow y_{u,t}}(S) = \mu_{y_{u,t+1}\rightarrow f_{u,t}}(S) \cdot B + \mu_{y_{u,t+1}\rightarrow f_{u,t}}(E_1) \cdot (A - B) \,. 
\end{equation}

\paragraph{Case $y_{u,t} = E_m, 1\leq m\leq M-1$} In this case, $f_{u,t}(\mathbf{y}^f_{u,t})$ is non-zero with value $1 - \pi(m;q_E)$ for $y_{u,t+1} = E_{m+1}$ and with value $\pi(m;q_E)$ for $y_{u,t+1} = I_1$, respectively. Also, $f_{u,t}(\mathbf{y}^f_{u,t})$ is independent of the infection states $y_{v_i,t}$ of contacts $v_i$ at time $t$. Thus, using (\ref{eq:A}) we get
\begin{equation*}
\mu_{f_{u,t}\rightarrow y_{u,t}}(E_m) = \left[\mu_{y_{u,t+1}\rightarrow f_{u,t}}(E_{m+1}) \cdot (1 - \pi(m;q_E)) + \mu_{y_{u,t+1}\rightarrow f_{u,t}}(I_1) \cdot \pi(m;q_E) \right] \cdot A \,.
\end{equation*}

\paragraph{Case $y_{u,t} = E_M$} In this case, $f_{u,t}(\mathbf{y}^f_{u,t})$ is non-zero with value $\pi(M;q_E)$ only for $y_{u,t+1} = I_1$ and is also independent of the infection states $y_{v_i,t}$ of contacts $v_i$ at time $t$. Thus, using (\ref{eq:A}) we get
\begin{equation}
\mu_{f_{u,t}\rightarrow y_{u,t}}(E_m) = \mu_{y_{u, t+1}\rightarrow f_{u,t}}(I_1) \cdot \pi(M;q_E) \cdot A \,.
\end{equation}

\paragraph{Case $y_{u,t} = I_n, 1\leq n\leq N-1$} In this case, $f_{u,t}(\mathbf{y}^f_{u,t})$ is non-zero with value $1-\pi(n;q_I)$ for $y_{u,t+1} = I_{n+1}$ and with value $\pi(n;q_I)$ for $y_{u,t+1} = R$, respectively. Also, $f_{u,t}(\mathbf{y}^f_{u,t})$ is independent of the infection states $y_{v_i,t}$ of contacts $v_i$ at time $t$. Thus, using (\ref{eq:A}) we get
\begin{equation*}
\mu_{f_{u,t}\rightarrow y_{u,t}}(I_n) = \left[\mu_{y_{u,t+1}\rightarrow f_{u,t}}(I_{n+1}) \cdot (1-\pi(n;q_I)) + \mu_{y_{u,t+1}\rightarrow f_{u,t}}(R) \cdot \pi(n;q_I)\right] \cdot A \,.
\end{equation*}

\paragraph{Case $y_{u,t} = I_N$} In this case, $f_{u,t}(\mathbf{y}^f_{u,t})$ is non-zero with value $\pi(N;q_I)$ only for $y_{u,t+1} = R$ and is also independent of the infection states $y_{v_i,t}$ of contacts $v_i$ at time $t$. Thus, using (\ref{eq:A}) we get 
\begin{equation}
\mu_{f_{u,t}\rightarrow y_{u,t}}(I_n) = \mu_{y_{u,t+1}\rightarrow f_{u,t}}(R) \cdot \pi(N;q_I) \cdot A \,.
\end{equation}

\paragraph{Case $y_{u,t} = R$} In this case, $f_{u,t}(\mathbf{y}^f_{u,t})$ is non-zero with value $1$ only for $y_{u,t+1} = R$ and is also independent of the infection states $y_{v_i,t}$ of contacts $v_i$ at time $t$. Thus, using (\ref{eq:A}) we get 
\begin{equation}
\mu_{f_{u,t}\rightarrow y_{u,t}}(R) = \mu_{y_{u,t+1} \rightarrow f_{u,t}}(R) \cdot A \,.
\end{equation}

\subsubsection{Message $\mu_{f_{u,t}\rightarrow y_{v_k,t}}$} 
In order to derive efficient message update equations, we note that in (\ref{eq:linear_reduction_template}), we no longer need to sum over the states of $y_{v_k,t}$ and thus define the following modifications of (\ref{eq:A}) and (\ref{eq:B})
\begin{eqnarray}
    A_k & := & \prod_{i \not= k} \left(\mu_{y_{v_i,t}\rightarrow f_{u,t}}(\mathcal{I}) + \mu_{y_{v_i,t}\rightarrow f_{u,t}}(\mathcal{N})\right) \label{eq:Ak} \,, \\
    B_k & := & (1-p_0) \cdot \prod_{i \not= k} \left(\mu_{y_{v_i,t}\rightarrow f_{u,t}}(\mathcal{I}) \cdot \prod_{j=1}^J (1-p_j)^{x_{ij}} + \mu_{y_{v_i,t}\rightarrow f_{u,t}}(\mathcal{N}) \right) \label{eq:Bk} \,.
\end{eqnarray}
We also note that the function $f_{u,t}(\mathbf{y}^f_{u,t})$ only changes values with respect to $y_{v_k,t}$ depending on $y_{v_k,t} \in \mathcal{I}$ or $y_{v_k,t} \in \mathcal{N}$. Thus, it is sufficient to only consider these two cases. 

\paragraph{Case $y_{v_k,t} \in \mathcal{I}$} In this case, there are $7$ combinations of $y_{u,t}$ and $y_{u,t+1}$ specified on the right-hand side of (\ref{eq:conditional_model_y}) that lead to non-zero values of the function $f_{u,t}(\mathbf{y}^f_{u,t})$
\begin{eqnarray}
    \mu_{f_{u,t} \rightarrow y_{v_k,t}}(y_{v_k,t}) & = 
    & \mu_{y_{u,t}\rightarrow f_{u,t}}(S) \cdot \mu_{y_{u,t+1}\rightarrow f_{u,t}}(S) \cdot \prod_{j=1}^J (1-p_j)^{x_{kj}} \cdot B_k + \nonumber \\
    & & \mu_{y_{u,t}\rightarrow f_{u,t}}(S) \cdot \mu_{y_{u,t+1}\rightarrow f_{u,t}}(E_1) \cdot \left(A_k - \prod_{j=1}^J (1-p_j)^{x_{kj}} \cdot B_k \right) + \nonumber \\
    & & \left(\sum_{m=1}^{M-1} \mu_{y_{u,t}\rightarrow f_{u,t}}(E_m) \cdot \mu_{y_{u,t+1}\rightarrow f_{u,t}}(E_{m+1}) \cdot (1-\pi(m;q_E)) \right) \cdot A_k + \nonumber \\
    & & \left(\sum_{m=1}^M \mu_{y_{u,t}\rightarrow f_{u,t}}(E_m) \cdot \mu_{y_{u,t+1}\rightarrow f_{u,t}}(I_1) \cdot \pi(m;q_E) \right) \cdot A_k + \nonumber \\
    & & \left(\sum_{n=1}^{N-1} \mu_{y_{u,t}\rightarrow f_{u,t}}(I_n) \cdot \mu_{y_{u,t+1}\rightarrow f_{u,t}}(I_{n+1}) \cdot (1-\pi(n;q_I)) \right) \cdot A_k + \nonumber \\
    & & \left(\sum_{n=1}^N \mu_{y_{u,t}\rightarrow f_{u,t}}(I_n) \cdot \mu_{y_{u,t+1}\rightarrow f_{u,t}}(R) \cdot \pi(n;q_I) \right) \cdot A_k + \nonumber \\
    & & \mu_{y_{u,t}\rightarrow f_{u,t}}(R) \cdot \mu_{y_{u,t+1}\rightarrow f_{u,t}}(R) \cdot A_k \,.
\end{eqnarray}
Note that the term $\prod_{j=1}^J (1-p_j)^{x_{kj}}$ has to be multiplied with $B_k$ because in this case, $y_{v_k,t} \in \mathcal{I}$.

\paragraph{Case $y_{v_k,t} \in \mathcal{N}$} In this case, there are also $7$ combinations of $y_{u,t}$ and $y_{u,t+1}$ specified on the right-hand side of (\ref{eq:conditional_model_y}) that lead to non-zero values of the function $f_{u,t}(\mathbf{y}^f_{u,t})$
\begin{eqnarray}
    \mu_{f_{u,t} \rightarrow y_{v_k,t}}(y_{v_k,t}) & = 
    & \mu_{y_{u,t}\rightarrow f_{u,t}}(S) \cdot \mu_{y_{u,t+1}\rightarrow f_{u,t}}(S) \cdot B_k + \nonumber \\
    & & \mu_{y_{u,t}\rightarrow f_{u,t}}(S) \cdot \mu_{y_{u,t+1}\rightarrow f_{u,t}}(E_1) \cdot \left(A_k - B_k \right) + \nonumber \\
    & & \left(\sum_{m=1}^{M-1} \mu_{y_{u,t}\rightarrow f_{u,t}}(E_m) \cdot \mu_{y_{u,t+1}\rightarrow f_{u,t}}(E_{m+1}) \cdot (1-\pi(m;q_E)) \right) \cdot A_k + \nonumber \\
    & & \left(\sum_{m=1}^M \mu_{y_{u,t}\rightarrow f_{u,t}}(E_m) \cdot \mu_{y_{u,t+1}\rightarrow f_{u,t}}(I_1) \cdot \pi(m;q_E) \right) \cdot A_k + \nonumber \\
    & & \left(\sum_{n=1}^{N-1} \mu_{y_{u,t}\rightarrow f_{u,t}}(I_n) \cdot \mu_{y_{u,t+1}\rightarrow f_{u,t}}(I_{n+1}) \cdot (1-\pi(n;q_I)) \right) \cdot A_k + \nonumber \\
    & & \left(\sum_{n=1}^N \mu_{y_{u,t}\rightarrow f_{u,t}}(I_n) \cdot \mu_{y_{u,t+1}\rightarrow f_{u,t}}(R) \cdot \pi(n;q_I) \right) \cdot A_k + \nonumber \\
    & & \mu_{y_{u,t}\rightarrow f_{u,t}}(R) \cdot \mu_{y_{u,t+1}\rightarrow f_{u,t}}(R) \cdot A_k \,.
\end{eqnarray}

\section{Simulation-Based Experimental Results}
In this section, we present two types of experimental evaluations:
\begin{enumerate}
    \item {\it Population Level COVID-19 Infection Spread}. In the first set of experiments, we will demonstrate that CRISP is capable of modelling infection spread across an entire population. We will relate our individual-level parameters ${\boldsymbol \theta}$ to more classical measures of infection spread such as reproduction factor $R_0$ and demonstrate that the structure of the contact patterns allow more fine grained control of the infection spread which can be used for alternative containment measures of the COVID-19 pandemic.
    \item {\it Test and Quarantine Efficacy of CRISP Model}. In the second set of experiments, we will assess the test and quarantine efficacy of the CRISP model by comparing the population health after 5 months under three testing and quarantining policies: (1) symptom-based, (2) contact-tracing-based, and (3) CRISP model-based.
\end{enumerate}
In all these experiments, we use the parameters $\alpha = 0.001$ and $\beta = 0.01$ in (\ref{eq:test_outcome}) and match the distribution $q_E$ and $q_I$ of exposure and infectiousness duration to the empirical distributions provided in the medical literature \cite{Backer2020,Woelfel2020}. This is both used in the generation of the simulated test outcome data as well as for the CRISP inference algorithms as these parameters are publicly known. We will use the notation $\overline{q_I}$ for the expectation of the empirical distributions $q_I$.

\begin{wrapfigure}{r}{0.35\textwidth}
    \includegraphics[width=0.35\textwidth]{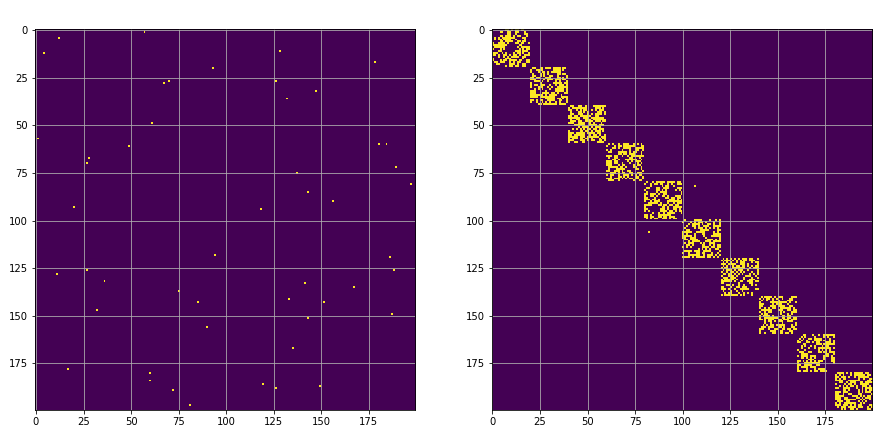}
    \caption{Snapshot of the contact matrix of the first 200 individuals for random connections and with "social bubbles". \label{fig:social_bubbles}}
\end{wrapfigure}
In order to simulate realistic epidemiological spread, we need to translate a reproduction factor $R_0$ at $t=0$ into contact data. By definition, $R_0$ is the average number of individuals that an infected person will infect over the entire period of being infectious. Thus, for a reproduction factor $R_0$ and a contact channel $j$ with transmission probability $p_j \in [0,1]$, we need to generate $C(R_0,p_j):=R_0/(\overline{q_I}\cdot p_j)$ many connections on average for all individuals in each time step.
Conversely, for any process that generates $\eta_j$ connections to unique and distinct individuals over channel $j$ in each time step, the effective $R_0$ over contact channel $j$ with 100\% transmission probability equals $\overline{q_I}\sum_j \eta_j$.
The actual number of contacts is drawn form a binomial distribution with $n=S-1$ and a rate $p=\frac{C(R_0,p_j)}{2(S-1)}$. Note that the rate is one half of the target contact rate because all contacts are symmetrically mirrored.

\subsection{Population Level COVID-19 Infection Spread} \label{subsec:population_experiments}
In order to assess if the CRISP model is able to provide realistic population-level statistics for COVID-19 infection spread, we simulate a population of $|\mathcal{S}|=10,000$ individuals over a period of 274 days (9 months). We single out an individual $u$ for whom we set $p_0=1$ so that she will get infected with probability 100\% at $t=1$ ("patient 0"); for all other people we assume a $p_0=10^{-6}$ to model a miniscule chance of infection spread from exogenous sources. We assume a single contact channel with a 1\% chance of transmission, $p_1=0.01$. We simulate five scenarios:
\begin{itemize}
    \item {\it No Mitigation}. Since $R_0$ of COVID-19 is estimated to be 2.5, at any time $t$ we generate $C(2.5,p_1)$ random connections for every individual at every time step.
    \item {\it Social Distancing After 60 Days}.  Intuitively, the "locality" of the contact patterns should play a role in the infection spread of COVID-19: if an individual is in contact with a broad range of other individuals, the spread should be faster than if unique number of people in contact over time is small. In order to demonstrate that this concept has indeed an effect on the infection spread, we performed an additional simulation where we kept the unique {\it number} of people that every individual meets in every time step at $C(2.5,p_1)$ but introduced the concept of "social bubbles" where all individuals form groups of 20 who have a large number of interactions with each other (i.e., equivalent to $C(2,p_1)$ but only rare interactions with people from other bubbles equivalent to $C(0.5,p_1)$ (see Figure \ref{fig:social_bubbles} for a picture of the contact matrix with random connections and with "social bubbles").
    \item {\it Mitigation After 60 Days}. For $t\leq60$, we generate $C(2.5,p_1)$ random connections for every individual at every time step. Afterwards, we assume that mitigation measures are taken which reduce the reproduction rate to $1.0$. Thus, we generate only $C(1.0,p_1)$ random connections for every individual at every time step $t>60$.
    \item {\it Suppression After 60 Days}. For $t\leq60$, we generate $C(2.5,p_1)$ connections for every individual at every time step. Afterwards, we assume that lock-down measures are taken to suppress the pandemic which reduce the reproduction rate to $0.5$. 
    \item{\it Suppression After 60 Days and Release of Lock-down after 120 Days}. This scenario is similar to the previous scenario but we assume that due to very low infection numbers, the lock-down is released after 60 days. Thus, we generate $C(2.5,p_1)$ random connections for every individual for $t > 120$.
\end{itemize}

\newcommand{\figtitles}[3]{\begin{minipage}{0.5\textwidth} \centering\textsf{\scriptsize #1}\\{#2}\\[-0.3em]\textsf{\tiny #3}\end{minipage}}
	
\begin{figure}
    	\figtitles 
    		{no mitigation}
    		{\includegraphics[width=\textwidth,trim={0 20 0 38},clip]{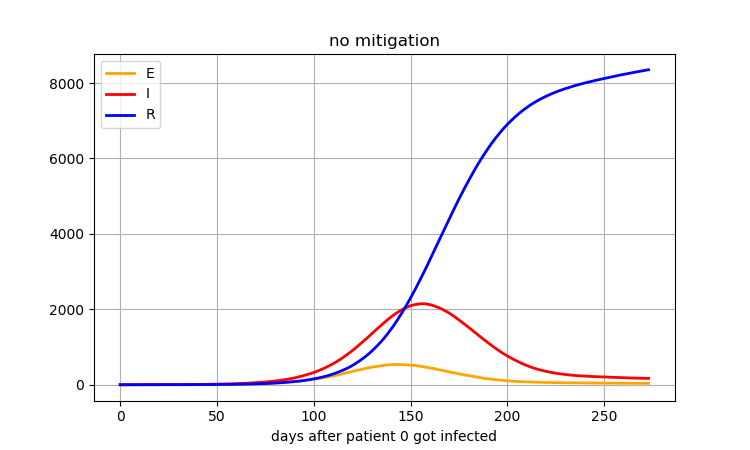}}
    		{days after patient 0 got infected}
    	\figtitles 
			{mitigation with localized contact pattern}
			{\includegraphics[width=\textwidth,trim={0 20 0 38},clip]{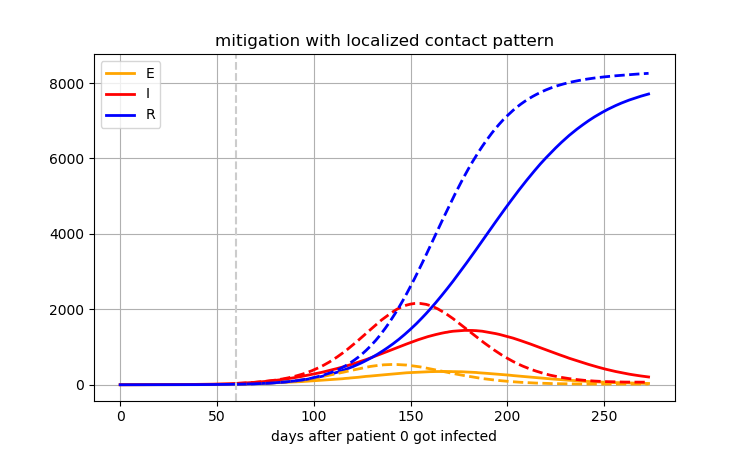}}
    		{days after patient 0 got infected}
    \\[0.5em]
    	\figtitles 
			{mitigation after 60 days}
			{\includegraphics[width=\textwidth,trim={0 20 0 38},clip]{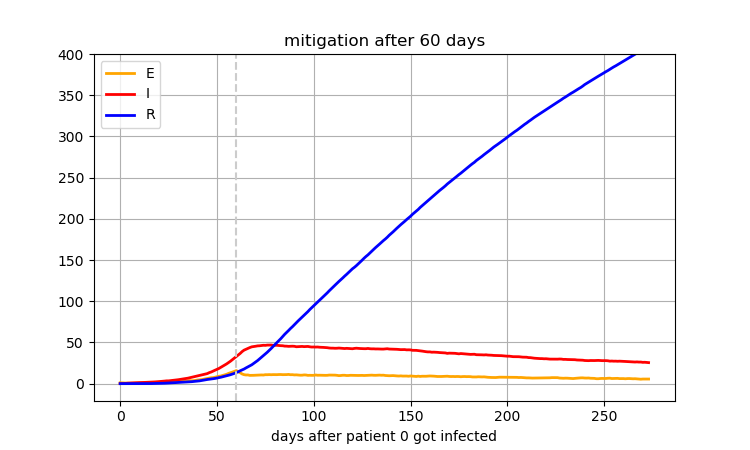}}
			{days after patient 0 got infected}
		\figtitles 
			{release after 60 days lock-down}
			{\includegraphics[width=\textwidth,trim={0 20 0 38},clip]{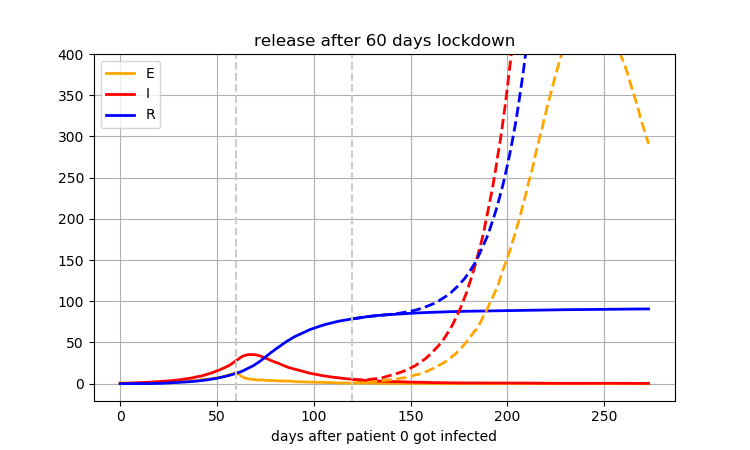}}
			{days after patient 0 got infected}
    \caption{Population level COVID-19 infection spread for three different scenarios: \textit{(top-left)} No mitigation ($R_0=2.5$). \textit{(top-right)} No mitigation until day 60 and then using "social bubbles". Note that $R_0$ remains at 2.5 the entire time. \textit{(bottom-left)} Mitigation after 60 days by reducing $R_0$ to $1.0$. \textit{(bottom-right)} Lock-down at day 60 and reduction of $R_0$ to $0.5$ (solid lines). In dashed lines we show the effect of a subsequent re-opening of a subsequent contact rate increase to $R_0$ of 2.5 starting at day 120. \label{fig:population}}
\end{figure}

In Figure \ref{fig:population}, we show the plot of $\sum_u P(z_{u,t}=z)$ over $t=1,\ldots,274$ days for $z\in\{E,I,R\}$ (orange = $E$, red = $I$, blue = $R$) from $100$ forward samples of the CRISP model for these scenarios. As one can see, with no mitigation there is a high peak around day $t^*=180$ and eventually herd-immunity is achieved at 85\% of infected population. Even though the number of unique contacts in each time step is the same, "social bubbles" flatten the curve, thus slowing down the infection but growth rates of infected people are still super linear until large parts of the population had been in contact with the disease. Note that a similar mitigation policy is currently used in Belgium.  In case of mitigation to $R_0=1.0$, growth rates are pushed to sub linear but the pandemic is still continuously going on after 9 months. Not surprisingly, suppression is most effective at bringing the infections back to nearly 0\% after 120 days. However, if the lock-down is lifted after $120$ days, a second wave of infections will cause an exponential increase in infectiousness after only two weeks (dashed lines). Note that all these effects were computable by simply forward sampling our individual-level CRISP model.

\subsection{Test and Quarantine Efficacy of CRISP Model} \label{subsec:quarantine_efficacy}
In order to assess the test and quarantine efficacy of the CRISP model, we consider a population of $|\mathcal{S}|=1,000$ individuals for $150$ days (5 months) with a uniformly random contact pattern of $C(2.5, 0.025)=5.03$ contacts on average per individual and day. We simulate the actual infection spread by applying the following sequence in each time step (i.e., day): At the beginning of each time step, we query the testing-and-quarantining policy for a list of individuals which need to be tested and need to be in quarantine during this step (this will only be done after $t^*=30$ to simulate an undetected initial outbreak). Each policy is constrained to select no more than $10$ test candidates per day (1\% of the total population). Given the quarantined individuals on that day, we remove contacts from and to the quarantined individuals for that day and then use the CRISP forward model (\ref{eq:conditional_model}) and CRISP test outcome model (\ref{eq:test_outcome}) to draw one sample of the next simulated infection state of every individual as well as the actual test outcomes of the requested test candidates. If the infection state of an individual changes from $E$ to $I$ in this sampling step, we assume that with 50\% probability, the individual generates symptoms. Finally, at the end of the time step, the testing-and-quarantining strategy is revealed the test outcomes as well as the list of symptomatic individuals (again, provided $t \geq t^*$). We single out an individual $u$ for whom we set $p_0=1$ so that she will get infected with probability 100\% at $t=1$ ("patient 0"); for all other people we assume a $p_0=10^{-4}$ to model a small chance of infection spread from exogenous sources.

\begin{enumerate}
	\item {\em Symptom-Based Policy}. For every time step $t \geq t^*$, we will request testing for up to 10 symptomatic individuals from the previous time step. For all individuals with a positive test outcome on the previous day, we will institute a quarantine for $\rho$ time steps where $\rho$ ranges from $2$ to $21$ days in our evaluation.
    \item {\em Contact-Tracing Policy}. For every time step $t \geq t^ *$, we will request testing for up to 10 symptomatic individuals from the previous time step. If there are less than 10 symptomatic individuals, then we will request the remaining tests for individuals in quarantine sorted in descending order of the number of contacts they have had in the past 7 days with  people who have tested positive. For every individual with a positive test outcome, we will not only quarantine her but also all the contacts she had in the past 7 days for $\rho$ time steps where $\rho$ ranges from $2$ to $21$ days in our evaluation; for every individual with a negative test outcome, we will remove her from quarantine.
    \item {\em CRISP Model-Based Policy}. For every time step $t \geq t^ *$, we will use block-Gibbs sampling of $100$ infection traces $\mathbf{z}_u$ to estimate $P(z_{u,t})$ for every individual $u$ at the current time step $t$ based on the contacts and test outcomes prior to time step $t$. We will request testing for up to 10 symptomatic individuals from the previous time step. If there are less than 10 symptomatic individuals, then we will request the remaining tests for individuals (who have not tested positive before) in descending order of $\hat{P}(z_{u,t}=I)$. We will quarantine any individual who is not yet quarantined but whose estimated probability $\hat{P}(z_{u,t}\in \{E,I\})$ exceeds a given policy threshold $\tau_{\mathrm{EI}}$; we will release an individual from quarantine once their estimated probability $\hat{P}(z_{u,t}\in \{S,R\})$ exceeds a given policy threshold $\tau_{\mathrm{SR}}$. Note that we increase $p_0$ in the block-Gibbs sampling by a factor of $10$ to account for "patient 0".
\end{enumerate}

In order to gauge the efficacy of each policy, we measure two quantities at the end of the simulation ($t=150$): (1) Percentage of population that got infected during the 150 days, and (2) total number of days that individuals were quarantined (e.g., if a policy locks down for the entire 150 days, this would result in 150,000 quarantine days). Varying the policy parameters $\rho$, $\tau_{\mathrm{EI}}$ and $\tau_{\mathrm{SR}}$ results in curves on the two dimensions of infection percentage and quarantine days. The closer a curve is to the origin, the more effective is the policy in terms of "health" (infection) and "economic" (quarantining) cost.

\begin{figure}[t]
    \begin{minipage}[m]{0.65\textwidth}
        \includegraphics[width=\columnwidth, trim={28 0 40 0}, clip]{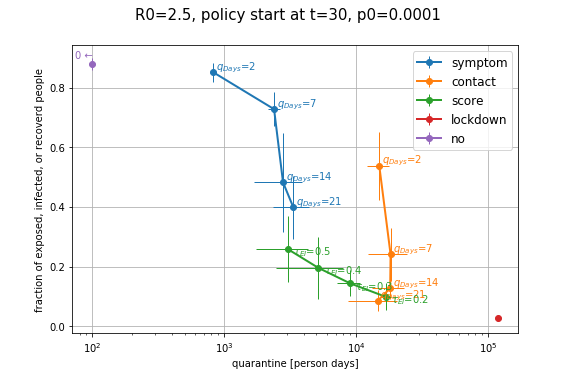}
    \end{minipage}
    \begin{minipage}[m]{0.35\textwidth}
        \caption{Effect of different mitigation policies on the infection percentage and quarantine days after $T=150$ days (5 months). The $y$-axis shows the percentage of population that got infected with COVID-19 during the 150 days. The $x$-axis shows the total number of days that individuals were quarantined. The error-bars are computed as the standard deviation over 20 random initializations of the forward model simulating the $T=150$ days while not affecting the randomization of the contact matrices.}
        \label{fig:exp_5.2}
    \end{minipage}
\end{figure}

\begin{figure}[ht!]
        \includegraphics[width=0.33\textwidth,trim={25 20 25 0},clip]{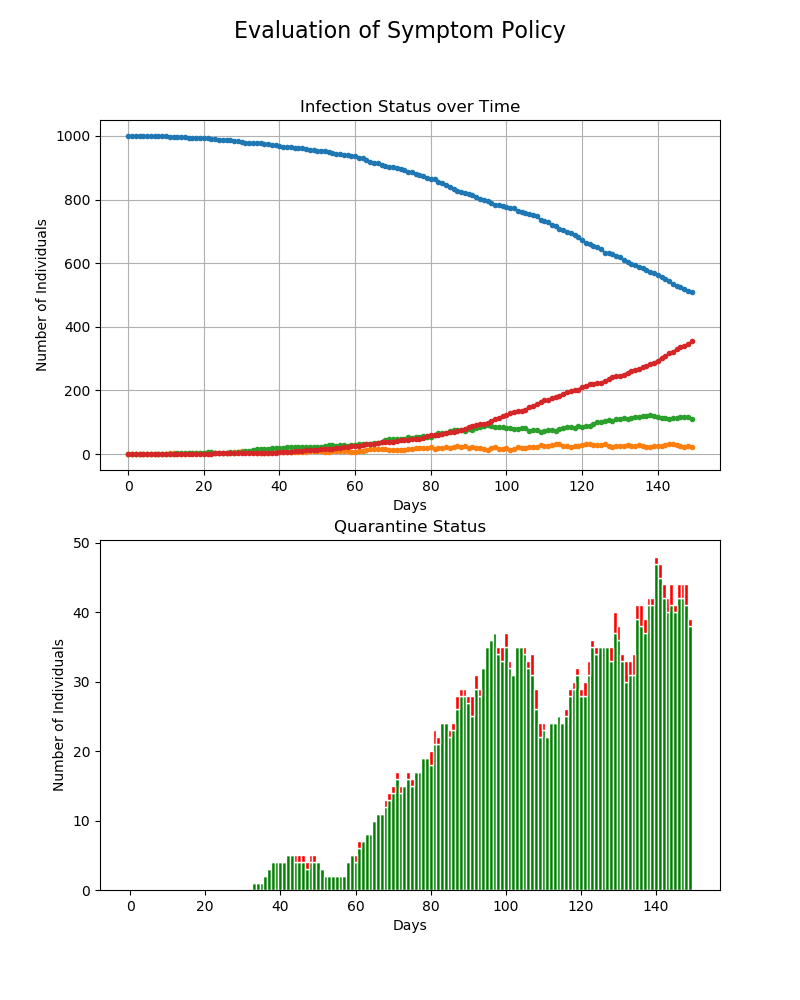}
        \includegraphics[width=0.33\textwidth,trim={25 20 25 0},clip]{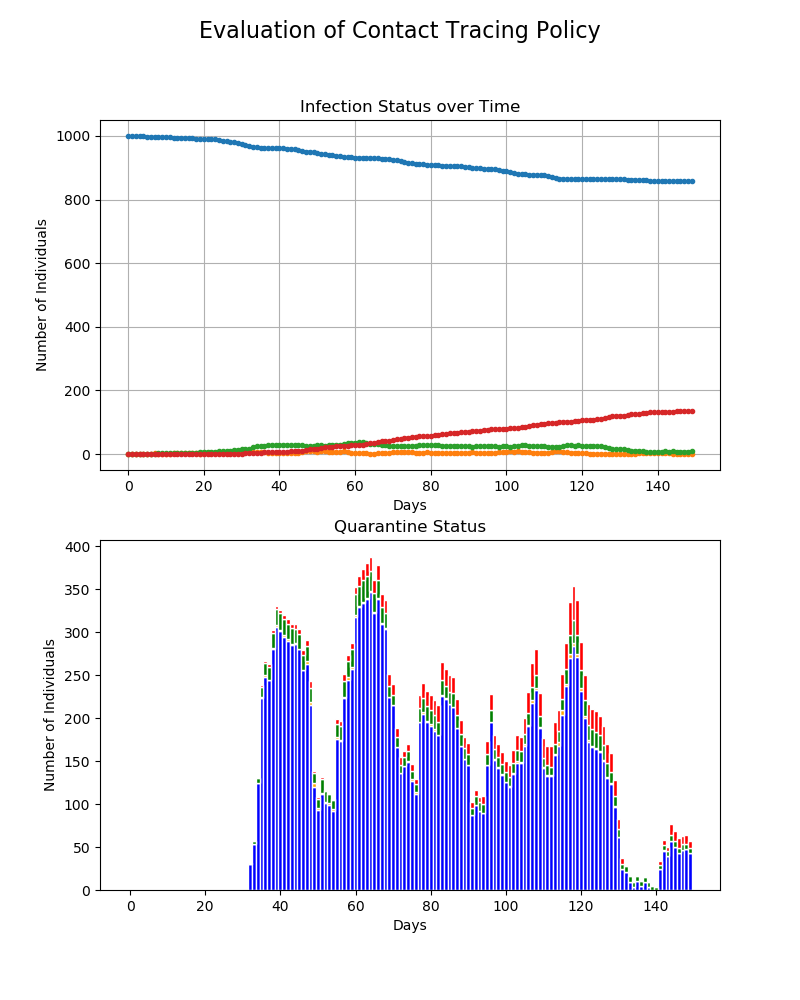} 
        \includegraphics[width=0.33\textwidth,trim={25 20 25 0},clip]{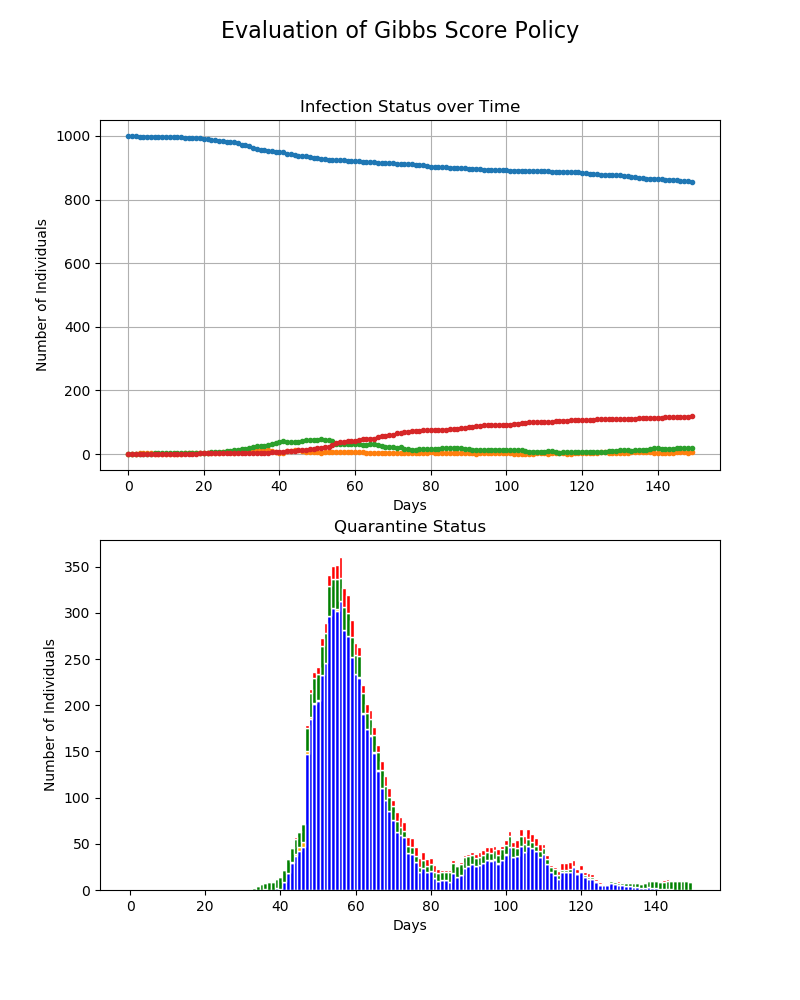} 
    \caption{Infection trace and quarantining statistics for symptom-based (left), contact-tracing (middle), and CRISP ($\tau_{\mathrm{EI}}=0.3$,$\tau_{\mathrm{SR}}=0.9$) model-based (right) testing-and-quarantining policy over the duration of 150 simulated days (blue = $S$, orange = $E$, green = $I$, red = $R$). In the bottom plots, we show a stacked bar chart of quarantined individuals per day grouped by actual infection status. While the number of quarantined individuals for the symptom-based policy is small, the infection spread is not contained and the quarantining keeps growing exponentially. In contrast, the contact-tracing policy effectively suppresses infection spread while regularly quarantining more than 25\% of the population. The CRISP model-based policy is initially picking a large number of individuals for quarantining but is then able to keep it at a low-level, in particular of susceptible individuals. \label{fig:evaluation}}
\end{figure}

In Figure~\ref{fig:exp_5.2}, we plot curves for the three policies with $\rho \in \{2,7,14,21\}$, $\tau_{\mathrm{EI}} \in \{0.2,0.3,0.4,0.5\}$, $\tau_{\mathrm{SR}} =0.9$. For comparison, we also show the two extreme points corresponding to "no mitigation" (i.e., zero quarantine days but the largest infection percentage of 90\%) and "full lock-down" (i.e., largest quarantine days of 120,000 and near-zero infection percentage). All three curves exhibit a negative slope where a higher percentage of quarantine days corresponds to a more effective mitigation of infection spread. Of the three policies, our CRISP-based policy achieves the best performance in terms of the smallest number of quarantine days for a given infection percentage (i.e., Pareto frontier). This is because our CRISP model is able to accurately identify infectious users (even though they may be asymptomatic) and test/quarantine them proactively– this helps to prevent infection from spreading across the population while at the same time quarantining fewer individuals with a high likelihood of getting infected. In contrast, the symptom-based policy only tests individuals with symptoms and then quarantines the individuals who have tested positive. As a result, since 50\% of the infected individuals are asymptomatic, they never get tested and quarantined, thus resulting in a spread of infection to 60\% of the population. Similarly, the contact-tracing policy, by isolating all contacts of positive tested individuals (many of whom may have low likelihoods of getting infected), is able to achieve the absolute smallest infection percentage but at the cost of massive quarantining (30\% of the population). Figure~\ref{fig:evaluation} shows a visualization of these effects in one of the simulation runs for $\rho = 14$, $\tau_{\mathrm{EI}}=0.3$, and $\tau_{\mathrm{SR}}=0.9$.

\section{Conclusions}
In this paper, we proposed a probabilistic graphical model for COVID-19 infection spread through individual contacts that captures the epidemiological state of each individual based on the SEIR model. We developed a computationally efficient block-Gibbs sampling-based algorithm to infer the COVID-19 infection risk score of all individuals at any time, given test outcome and mutual contact information between individuals. An efficient C++-based Python implementation of our inference algorithm is available at \href{https://github.com/zalandoresearch/CRISP}{https://github.com/zalandoresearch/CRISP}. Through experiments with simulated data, we showed that the CRISP model is able to model macro-level characteristics of the COVID-19 infection at county level ($\approx 10,000$ individuals) and effectively mitigate COVID-19 spread by pro-actively quarantining and testing individuals with high risk of infections.

As part of future work, we would like to further accelerate our inference procedure using other approximation techniques such as Variational Bayes \cite{Bishop2006}. Our inference algorithm can also be speeded up by exploiting the parallelism inherent in our block-Gibbs Sampling algorithm. For example, it is possible to concurrently sample infection traces of two individuals with no contacts in common. It is also known that the hyper-parameters of the SEIR model vary with demographic attributes such as age, socio-economic status, or location (see, for example \cite{Lorch2020} who present a location-varying infection spread model). We would like to extend our model with group-level hyper-parameters to account for this variation. We would also like to explore the causal impact of mitigation or suppression policy measures (e.g., school closures, shop closures, small group gatherings) on COVID-19 infection spread when using contact-level data. Finally, we would like to consider more sophisticated models of COVID-19 transmission through different modalities, and contacts with varying duration and distance characteristics.

\paragraph{Acknowledgments}
We would like to thank Sebastian Munoz, Christopher Gandrud, Jasvinder Kandola and Peter Herbrich for all their valuable input and feedback on earlier drafts of this paper. We are also indebted to Christoph Thöns for his support with Python and C++ programming. 

\appendix
\section{Proofs}

\begin{lem} \label{lem:expansion}
    For any $\kappa \in \mathbb{R}$ and any sequence of $n$ numbers $a_i \in \mathbb{R}$ and $b_i \in \mathbb{R}$,
    \begin{equation}
        \kappa \cdot \sum_{q_1 \in \{0, 1\}} \cdots \sum_{q_n \in \{0, 1\}} \prod_{i=1}^n \left(a_i^{q_i} \cdot b_i^{1-q_i}\right) = \kappa \cdot \prod_{i=1}^n (a_i + b_i) \,.
    \end{equation}
\end{lem}
\begin{proof}
    Expanding each summand directly, we see that exactly one of $a_i$ or $b_i$ is active, but never both. Thus,
    \begin{eqnarray*}
        & & \kappa \cdot \sum_{q_1 \in \{0, 1\}} \cdots \sum_{q_n \in \{0, 1\}} \prod_{i=1}^n \left(a_i^{q_i} \cdot b_i^{1-q_i}\right) \\ 
        & = & \kappa \cdot \sum_{q_1 \in \{0, 1\}} \left(a_1^{q_1} \cdot b_1^{1-q_1}\right) \cdot \sum_{q_2 \in \{0, 1\}} \cdots \sum_{q_n = 0, 1} \prod_{i=2}^{n} \left(a_i^{q_i} \cdot b_i^{1-q_i}\right) \\
        & = & \kappa\cdot \left(a_1\cdot \sum_{q_2 \in \{0, 1\}} \cdots \sum_{q_n \in \{0, 1\}} \prod_{i=2}^{n} \left(a_i^{q_i} \cdot b_i^{1-q_i}\right) + b_1\cdot \sum_{q_2 \in \{0, 1\}} \cdots \sum_{q_n \in \{0, 1\}} \prod_{i=2}^{n} \left(a_i^{q_i}   \cdot b_i^{1-q_i}\right)\right)\\
        & = & \kappa\cdot (a_1 + b_1) \cdot \sum_{q_2 \in \{0, 1\}} \cdots \sum_{q_n \in \{0, 1\}} \prod_{i=2}^{n} \left(a_i^{q_i} \cdot b_i^{1-q_i}\right) \\
        & = & \kappa \cdot \prod_{i=1}^n (a_i + b_i)
    \end{eqnarray*}
\end{proof}

\bibliographystyle{plain}
\bibliography{crisp}

\begin{thebibliography}{10}

\bibitem{Backer2020}
Jantien~A Backer, Don Klinkenberg, and Jacco Wallinga.
\newblock Incubation period of 2019 novel coronavirus (2019-{nCoV}) infections
  among travellers from {W}uhan, {C}hina, 20-28 {J}anuary 2020.
\newblock {\em Euro Surveillance}, 25(5), 2020.

\bibitem{Bishop2006}
Christopher Bishop.
\newblock {\em Pattern Recognition and Machine Learning}.
\newblock Springer, 2006.

\bibitem{Chakrabarti2008}
D.~Chakrabarti, Y.~Wang, C.~Wang, J.~Leskovec, and C.~Faloutsos.
\newblock Epidemic thresholds in real networks.
\newblock {\em ACM Transactions on Information and System Security}, 10(4),
  2008.

\bibitem{Deardon2010}
R.~Deardon, S.~P. Brooks, B.~T. Grenfell, M.~J. Keeling, M.~J. Tildesley, N.~J.
  Savill, D.~J. Shaw, and M.~E. Woolhouse.
\newblock Inference for individual-level models of infectious diseases in large
  populations.
\newblock {\em Statistica Sinica}, 20(1), 2010.

\bibitem{Ferguson2006}
N.~M. Ferguson, D.~A. Cummings, C.~Fraser, J.~C. Cajka, P.~C. Colley, and D.~S.
  Burke.
\newblock Strategies for mitigating an influence pandemic.
\newblock {\em Nature}, 442(7101):448--452, 2006.

\bibitem{Ferguson2005}
N.~M. Ferguson, D.A. Cummings, S.~Cauchemez, C.~Fraser, S.~Riley, A.~Meeyai,
  S.~Iamsirithaworn, and D.~S. Burke.
\newblock Strategies for containing an emerging influenza pandemic in southeast
  asia.
\newblock {\em Nature}, 437(7056):209--214, 2005.

\bibitem{Goyal2010}
Amit Goyal, Francesco Bonchi, and Laks Lakshmanan.
\newblock Learning influence probabilities in social networks.
\newblock In {\em Web Search and Data Mining (WSDM)}, 2010.

\bibitem{Kempe2003}
D.~Kempe, J.~Kleinberg, and E.~Tardos.
\newblock Maximizing the spread of influence through a social network.
\newblock In {\em International Colloquium on Automata, Languages and
  Programming}, 2003.

\bibitem{KscFreLoe2001}
Frank~R. Kschischang, Brendan~J. Frey, and Hans-Andrea Loeliger.
\newblock Factor graphs and the sum-product algorithm.
\newblock {\em {IEEE} Transaction on Information Theory}, 47(2):498--519, 2001.

\bibitem{Yuchen2018}
Yuchen Li, Ju~Fan, Yanhao Wang, and Kian-Lee Tan.
\newblock Influence maximization on social graphs.
\newblock {\em IEEE Transactions on Knwledge and Data Engineering},
  30(10):1852--1872, 2018.

\bibitem{Lorch2020}
Lars Lorch, William Trouleau, Stratis Tsirtsis, Bernhard Sch{\"o}lkopf, and
  Manuel Gomez-Rodriguez.
\newblock A spatiotemporal epidemic model to quantify the effects of contact
  tracing, testing, and containment.
\newblock {\em arXiv:2004.07641v2}, 2020.

\bibitem{Mathioudakis2011}
Michael Mathioudakis, Francesco Bonchi, Carlos Castillo, Aristides Gionis, and
  Antti Ukkonen.
\newblock Sparsification of influence networks.
\newblock In {\em International Conference on Knowledge Discovery and Data
  Mining}, 2011.

\bibitem{May1991}
Robert~M. May.
\newblock {\em Infectious diseases of humans: dynamics and control}.
\newblock Oxford University Press, 1991.

\bibitem{McMahan2017}
H.B. McMahan, E.~Moore, D.~Ramage, S.~Hampson, and B.A. y~Arcas.
\newblock Communication-efficient learning of deep networks from decentralized
  data.
\newblock In {\em Proceedings of the 20th International Conference on
  Artificial Intelligence and Statistics}, pages 1273--1282, 2017.

\bibitem{Mieghem2014}
P.~Van Mieghem and J.~Omic.
\newblock In-homogeneous virus spread in networks.
\newblock {\em arXiv:1306.2588v2}, 2014.

\bibitem{Mieghem2009}
P.~Van Mieghem, J.~Omic, and R.~Kooij.
\newblock Virus spread in networks.
\newblock {\em IEEE/ACM Transactions on Networks}, 2009.

\bibitem{Myers2010}
Seth Myers and Jure Leskovec.
\newblock On the convexity of latent social network inference.
\newblock In {\em Neural Information Processing Systems}, 2010.

\bibitem{Warriyar2020}
Vineetha Warriyar K.~V., Waleed Almutiry, and Rob Deardon.
\newblock Individual-level modeling of infectious disease data: Epiilm.
\newblock {\em arXiv:2003.04963v1}, 2020.

\bibitem{Woelfel2020}
Roman Woelfel, Victor~Max Corman, Wolfgang Guggemos, Michael Seilmaier, Sabine
  Zange, Marcel~A Mueller, Daniela Niemeyer, Patrick Vollmar, Camilla Rothe,
  Michael Hoelscher, Tobias Bleicker, Sebastian Bruenink, Julia Schneider,
  Rosina Ehmann, Katrin Zwirglmaier, Christian Drosten, and Clemens Wendtner.
\newblock Clinical presentation and virological assessment of hospitalized
  cases of coronavirus disease 2019 in a travel-associated transmission
  cluster.
\newblock {\em medrxiv.org:10.1101/2020.03.05.20030502v1}, 2020.

\bibitem{AarogyaSetu}
WWW.
\newblock Aarogya setu.
\newblock \url{https://www.mygov.in/aarogya-setu-app/}, 2020.
\newblock Accessed: 2020-05-10.

\bibitem{TraceTogether}
WWW.
\newblock Trace together.
\newblock \url{https://www.tracetogether.gov.sg}, 2020.
\newblock Accessed: 2020-05-10.

\end{thebibliography}

\end{document}